\documentclass[twocolumn]{aastex61}
\usepackage{subfigure}
\usepackage{amsmath}
\newcommand{\ups}{$\Upsilon$}
\newcommand{\msun}{$M_\odot$}
\shortauthors{Parsa et. al}

\begin{document}

\title{Investigating the Relativistic Motion of the Stars Near the Supermassive Black Hole in the Galactic Center}


\author{M. Parsa}
\affiliation{I. Physikalisches Institut, Universit\"at zu K\"oln, Z\"ulpicher Str. 77, 50937 K\"oln, Germany}
\affiliation{Max-Planck-Institut f\"ur Radioastronomie, Auf dem H\"ugel 69, 53121 Bonn, Germany}

\author{A. Eckart}
\affiliation{I. Physikalisches Institut, Universit\"at zu K\"oln, Z\"ulpicher Str. 77, 50937 K\"oln, Germany}
\affiliation{Max-Planck-Institut f\"ur Radioastronomie, Auf dem H\"ugel 69, 53121 Bonn, Germany}

\author{B. Shahzamanian}
\affiliation{I. Physikalisches Institut, Universit\"at zu K\"oln, Z\"ulpicher Str. 77, 50937 K\"oln, Germany}

\author{V. Karas}
\affiliation{Astronomical Institute, Academy of Science, Bo\v{c}n\'{i} II 1401, CZ-14131 Prague, Czech Republic}

\author{M. Zaja\v{c}ek}
\affiliation{I. Physikalisches Institut, Universit\"at zu K\"oln, Z\"ulpicher Str. 77, 50937 K\"oln, Germany}
\affiliation{Max-Planck-Institut f\"ur Radioastronomie, Auf dem H\"ugel 69, 53121 Bonn, Germany}

\author{J. A. Zensus}
\affiliation{Max-Planck-Institut f\"ur Radioastronomie, Auf dem H\"ugel 69, 53121 Bonn, Germany}

\author{C. Straubmeier}
\affiliation{I. Physikalisches Institut, Universit\"at zu K\"oln, Z\"ulpicher Str. 77, 50937 K\"oln, Germany}

\begin{abstract}

The S-star cluster in the Galactic center allows us to study the physics close to a supermassive black hole, including distinctive dynamical tests of general relativity. Our best estimates for the mass of and the distance to Sgr A* using the three stars with the shortest period (S2, S38, and S55/S0-102) and Newtonian models are $M_{BH} = (4.15 \pm 0.13 \pm 0.57) \times 10^{6}$~\msun\ and $R_0 = 8.19 \pm 0.11 \pm 0.34$~kpc. Additionally, we aim at a new and practical method to investigate the relativistic orbits of stars in the gravitational field near Sgr A*. We use a first-order post-Newtonian approximation to calculate the stellar orbits with a broad range of periapse distance $r_p$. We present a method that employs the changes in orbital elements derived from elliptical fits to different sections of the orbit. These changes are correlated with the relativistic parameter defined as \ups~$\equiv~r_s/r_p$ (with $r_s$ being the Schwarzschild radius) and can be used to derive \ups\ from observational data. For S2 we find a value of \ups~= 0.00088 $\pm$ 0.00080, which is consistent, within the uncertainty, with the expected value of \ups~= 0.00065 derived from $M_{BH}$ and the orbit of S2. We argue that the derived quantity is unlikely to be dominated by perturbing influences such as noise on the derived stellar positions, field rotation, and drifts in black hole mass.

\end{abstract}

\keywords{Galaxy: center --- astrometry --- infrared: stars -- black hole physics}

\section{Introduction}

\indent Monitoring the small group of fast-moving stars in the Galactic center (GC), known as the S-stars \citep{Eckart97}, 
has uncovered the existence of a $ \sim 4 \times 10^6$~\msun\ supermassive black hole (SMBH), Sagittarius~A* (Sgr~A*), 
located in the central stellar cluster of the Milky Way \citep[e.g.][]{Eckart1996, Ghez1998, Eckart2017}.
The small distance to Sgr~A* and the high velocities of some of the S-stars during their periapse passage 
have triggered the investigations to test 
the predictions of general relativity (GR) in the vicinity of the black hole
\citep{jaro98, f&m99, rubilar2001, weinberg5, zuc2006}.
However, tests of GR depend on an accurate knowledge of the gravitational potential, requiring precise observations of stellar orbits and meticulous determination
of the mass of the SMBH ($M_{BH}$) and the distance to it ($R_0$).

\indent The mass of Sgr~A* and the distance to the GC
are important quantities.
They allow us to  place the Milky Way in the observed
correlations between the central black hole mass, the
velocity dispersion, and the luminosity of the bulge stars
\citep{ferrarese2002, Tremaine2002, kormendy2013}.
The quantity $R_{0}$ is an essential basis for the understanding and modeling the Galactic dynamics \citep[e.g.][]{EnglmaierGerhard1999,Portail2016}.
The two quantities together allow us to determine the apparent size of the Schwarzschild radius on the sky.
If Sgr~A* has a spin and a suitable orientation, and if the luminous accretion zone is not heavily disturbed,
one can expect to see the shadow of a black hole--the size of which depends on $R_{0}$ and $M_{BH}$
\citep[e.g.][]{Falcke2000,FragaEncinas2016}.

\indent {\it Stars close to Sgr~A*.} One of the brightest stars (near-infrared (NIR) $K_s$-band (centered at 2.18~$\mu m$ with a width of 0.35~$\mu m$) magnitude = 14.2) in this cluster is S2 (also referred to as S0-2). 
While the first Keplerian orbital elements of the S2's orbit could be derived from the pre-periapse data, 
the situation improved significantly after the periapse passage of S2 in 2002.
NIR adaptive optics (AO) imaging allowed the derivation of 
detailed Keplerian orbital elements \citep{schodel2002, ghez3}.
NIR spectroscopy resulted in radial velocities, and hence in a determination of the mass and the distance
to the GC derived from the orbital data
\citep[$R_0$;][]{ghez3, ghez5, eisenhauer2003}.
These results could be improved using other high-velocity S-stars in the vicinity of Sgr~A*
\citep[e.g.][]{ghez8, gill2009, boehle16, gill17}.  
So far the orbits of 40 of these stars at distances between 1 and 100~milliparsecs (mpc) from Sgr~A* have been determined 
using NIR imaging and spectroscopy \citep{gill17}. 
However, it is presumed that there are many more fainter stars in the innermost region around Sgr~A* to be discovered
 \citep{sabha12}.
The star S2 has a short orbital period of about 16.2~yr which has enabled us to observe it long enough to 
determine its motion with high accuracy \citep{ghez3, gill2009, gill17}. 
Its velocity at periapse in units of the speed of light is $\sim$0.02, and if we consider a semimajor axis of 0\arcsec.124, an eccentricity of 0.88, and a BH mass of 4 $\times$ 10$^6$~\msun, then the periapse shift to the lowest order 
will be around 10$'$.8 which is large enough to be 
observable \citep{jaro98, f&m99, rubilar2001, weinberg5}.

\indent The proper motion of S38 ($K_s$=17; also referred to as S0-38) can also help us in determining the gravitational potential parameters of the SMBH with high accuracy, since a large portion of its orbit has been observed and the rest can be covered in a short time. Another reason that makes S38 important is that although the orbit is located in the very central region of the GC, most of its orbit is to the west of Sgr~A* on the sky, which is much less crowded than the other parts of the center, and thus S38 is less prone to confusion with other sources in the center. Moreover, the large uncertainty in determining the closest approach of S2 to the SMBH has limited us in determining the north--south motion of the black hole, while the fact that the orbit of S38 is perpendicular to the orbit of S2 can help us in overcoming this limit \citep{boehle16}. The orbit of S38 has been used combined with the orbit of S2 to constrain the gravitational potential in the GC in \citet{boehle16}.

\indent A further important source is S55/S0-102 \citep[$K_s$=17.1;][]{meyer12}
 with a period of just $\sim$12 yr, which makes it the star with the shortest period yet known. If stars within the orbit of S2, S38, and S55/S0-102 are discovered, spin--related effects, e.g., the Lense--Thirring precession and the frame-dragging effect, and therefore tests of the no-hair theorem, appear to be in reach \citep{preto2009, ang&saha10, merritt10, zhang15}.

\indent {\it Post-Newtonian effects.} Shortly after GR had been formulated (by Einstein in 1915)\footnote{For recent overviews on its centenary, see, e.g., \citet{iorio15} and \citet{debono16}.} it was recognized that three most promising observational tests can be set up in the regime of the weak gravitational field of the Solar system: (i) the measurement of the deflection of light passing close to a gravitating body, (ii) the time delay of light traversing the gravitational field, and (iii) the shift of the pericenter of an orbit of a test body on a closed trajectory (exhibited as the anomalous perihelion shift of Mercury).

\indent The experimental constraints on the shift of Mercury's perihelion sparked confusion because the total value of the shift contains additional influences, and it was also realized that these measurements could be improved if perihelion shifts from different planets (orbiting at different radii) are included.

\indent In our present work we also adopt the shift of the pericenter as a suitable and practical approach to check the character of the gravitational field near Sgr~A*. Several S-cluster stars can be employed as test particles on different orbits around the central black hole, hence improving the precision. As a consequence of our setup, we can tackle the problem within the framework of the weak-field post-Newtonian (PN) approximation. A source of complication (similar to the historical case of Mercury) is caused by a potential role of the mass of up to 10$^4$ solar masses \citep[e.g. the robust early result by][]{Mouawad2005}
 distributed within the orbit of stars, which may cause a 
Newtonian precession of the same order of magnitude as GR precession but in the opposite direction. \citet{rubilar2001} studied this effect and concluded that this Newtonian shift may partially or 
completely compensate the GR shift for S2-like stars. Moreover, the granularity of the distributed mass 
(i.e. the number of the perturbers) may affect both the eccentricity and the orbital plane through the phenomenon of resonant 
relaxation \citep{sabha12}. These perturbations possibly hide the frame-dragging and the 
Lense--Thirring effects for the orbits larger than $\sim$1~mpc and $\sim$0.2~mpc, respectively \citep{merritt10}.

\indent Determining the proper motions and the radial velocities of the stars can lead to the detection of manifestations of these perturbations in curves of stellar redshift, i.e., the gravitational redshift and the special relativistic transverse Doppler effect \citep{zuc2006, ang&saha10, zhang15}. This might be observable during the next periapse passage of S2 in $\sim$ 2018.6. \citet{zuc2006} conclude the detection of the PN effects in the redshift curves will be possible only after a decade of observation. \citet{iorio&zhang} investigate the possibility of using the PN corrections of the Schwarzschild-like, Lense--Thirring, and quadrupole momentum effects to the orbital period in order to conduct new tests of GR. \citet{zhang&iorio} study the gravitational perturbations on the apparent position on the sky plane and the redshift of the stars and stellar remnants around the BH in order to investigate the possibility of unbiased measurements of spin-induced effects of a Kerr BH. They also study the possible perturbations from S55/S0-102 on the orbit of S2. \citet{hees17} constrain a fifth force using Keck observations of the two short-period stars in search of deviations from GR. Moreover, they put an upper limit on a shift of the argument of periapse produced by a fifth force that is one order of magnitude larger than the periapse shift due to GR. The relativistic effects are stronger and probably dominant in the stars with smaller orbits (shorter periods) than S2, which makes the detection of the perturbation due to Gr more promising. The discovery of such stars is highly anticipated in the near future using instruments such as the GRAVITY on the Very Large Telescope Interferometer 
\citep[VLTI, e.g.][]{Eisenhauer2011, Eckart2012, Grould2017} which is currently being commissioned and the European Extremely Large Telescope 
\citep[E-ELT, e.g.][]{Brandl2016, Davies2016} which is under develop.

\indent In this work, we use the NIR imaging and spectroscopy data of the three known stars closest to Sgr A* to study the properties of the central black hole, i.e. mass and the distance to it. We investigate the PN effects observable in the orbital motion of the S2 star with the smallest known periapse distance as well as simulated stars within its orbit. We start with the details of the observations in NIR, the data reductions, and the Keplerian and relativistic PN models in Sect. \ref{sec:obs}. We discuss our astrometric accuracy and find the astrometric positions of our candidate S-stars (S2, S38, and S55/S0-102) to derive the best orbital fits, both Keplerian and relativistic, and consequently obtain the gravitational potential parameters in Sect. \ref{sec:orbits}. In Sect. \ref{sec:cases} we develop two methods that use the deviations from a Newtonian symmetric orbit in two directions to measure PN effects within the weak-field limit. Since S2 is the only S-star with a small enough periapse distance that makes the observation of these effects promising and it is the only star with enough available data on astrometric and radial velocities, it is necessary to start by simulating the orbits of the stars that are located within the orbit of S2 as our highly to mildly relativistic case studies. The results from these simulations are analyzed by connecting them to the relativistic parameter at periapse, which is beneficial in assessing the magnitude of PN effects. The relativistic parameter at periapse is correlated with the mass of the BH, the periapse distance, the relativistic periapse precession, and the relativistic $\beta$. The results are then applied to S2 in order to evaluate and confirm the effectiveness of our methods and are presented in Sect. \ref{sec:results}. The results from all sections are discussed in Sect. \ref{sec:comb} and finally a summary of the conclusions is given in Sect. \ref{sec:con}.

\section{Observations and Simulations}
\label{sec:obs}

\indent We can observe a large portion of the orbits of stars with short orbital periods to enable us to determine their motion with precision and thus to study the properties of our Galaxy's SMBH. Moreover, the stars with small orbits and small closest approach to the SMBH of these stars make them the best candidates for investigating the effects of GR. Therefore, we choose S2, S38, and S55/S0-102 as our candidate S-stars. Making use of the previously reported astrometric and radial velocity data of these stars will also help us to cover more of their orbits.

\indent The first step is to obtain the astrometric positions from the near-infrared data to find a Newtonian model for their motions. Additionally we need a model to describe their relativistic motion around the SMBH for tests of GR.

\subsection{Near-infrared Data}
\label{sec:data}

\indent The positions of the S-stars are calculated from the AO-assisted imaging data of the GC from 2002 to 2015 taken by the NAOS-CONICA (NACO) instrument installed at the fourth (from 2001 to 2013) and then the first (from 2014 on) unit telescope of the Very Large Telescope (VLT). The $K_s$-band (2.18~$\mu$m) images obtained by the S13 camera (with 13~mas~pix$^{-1}$ scale) and the S27 camera of NACO (with 27~mas~pix$^{-1}$ scale) are used. The AO guide star is IRS7 with $K_s$~= 6.5--7.0~mag located at about 5$\arcsec$.5 north of Sgr~A*. The data reduction consists of the standard steps--flat-fielding, sky subtraction, and bad-pixel correction. A cross-correlation algorithm is used to align the dithered exposures. We use the 27~mas~pix$^{-1}$ scale images to measure the position of the SiO maser stars IRS9, IRS10EE, IRS12N, IRS15NE, IRS17, IRS19NW, IRS28, and SiO-15 \citep{Menten1997,Reid2003, reid7} which were needed for finding the connection of the NACO NIR data and the radio reference frame. In order to measure the position of the S-stars, the Lucy--Richardson deconvolution algorithm is used to resolve the sources in the 13~mas~pix$^{-1}$ scale images. For each epoch we included all available $K_s$-band frames of the GC stellar cluster that were taken with a close to diffraction-limited AO correction and showed Sgr~A* flaring. We use the reduced data presented by \citet[][Table 2]{witzel12}, 2003 to mid-2010, and \citet[][Table 1]{eckart13} and \citet[][Table 1]{shah15}, 2002--2012. For the remaining years, 2013--2015, the positions are obtained by observing flare activity of the black hole (see Table~\ref{table:obstable}\footnote{ProgramIDs: 60.A-9026(A), 713-0078(A), 073.B-0775(A), 073.B-0085(E), 073.B-0085(F), 077.B-0552(A), 273.B.5023(C), 073-B-0085(I), 077.B-0014(C), 077.B-0014(D), 077.B-0014(F), 078.B-0136(A), 179.B-0261(A), 179.B-0261(H), 179.B-0261(L), 179.B-0261(M), 179.B-0261(T), 179.B-0261(N), 179.B-0261(U), 178.B-0261(W), 183.B-0100(G), 183.B-0100(D), 183.B-0100(I), 183.B-0100(J), 183.B-0100(T), 183.B-0100(U), 183.B-0100(V), 087.B-0017(A), 089.B-0145(A), 091.B-0183(A), 095.B-0003(A), 081.B-0648(A), 091.B-0172(A)}). 
The radial velocity data used for S2 are from the AO-assisted field spectrometer SINFONI installed on the fourth unit telescope of the VLT and taken from \citet{gill2009}. The radial velocity measurements used for S38 are from \citet{boehle16}.
The orbital fits presented in section~\ref{sec:orbits} were all exclusively 
done with the VLT stellar positions and the radial velocities as mentioned 
above. However, when discussing methods to derive the relativistic parameter and in 
particular differences in the argument of the periapse $\omega$ starting in 
chapter section~\ref{sec:method},
we used in addition, for the stars S2 and S38, the positions published by
\citet{boehle16} for the years 1995--2010 and 2004--2013, respectively.

\begin{table}[htbp]
\centering
\caption{Summary of observations used in addition to \citet[][Table 2]{witzel12}, and \citet[][Table 1]{eckart13}, and \citet[][Table 1]{shah15} from 2013 to 2015.}
\label{table:obstable}
\begin{tabular}{l c c}
\hline\hline
\hspace{1.25cm} Date & & Camera\\
\hspace{0.5cm}(UT) & (Decimal) & \\
\hline
 2013 Jun 5 & 2013.425 & S27\\
 2013 Jun 28& 2013.488 & S13\\
 2015 Aug 1  & 2015.581 & S13\\
\hline
\end{tabular}
\end{table}

\subsection{Simulations}
\label{sec:simul}

\indent To investigate the effects of GR and measure their strength on some of the S-stars and the stars located within the orbit of S2, one should use a model for their relativistic non-Newtonian orbits. Here we used thePN approximation given in \cite{einstein} known as the Einstein--Infeld--Hoffmann equations of motion. The PN approximation (see \cite{Wei72,Will93}, also \cite{Sch96}) applies to the particles that are bound in a gravitational field and have small velocities with respect to the  velocity of light. It is based on an expansion of the quantities that determine the particle trajectory. Rewriting the equation for the gravitational potential $\phi=-{GM_{BH}}/{r}$ of a compact mass distribution of a total mass $M_{BH}$ and allowing it to move with a constant velocity, we can write the equation of motion of a star as

\begin{equation}
\label{eqn:eih}
\begin{split}
\frac{d\pmb{v}_{\star}}{dt} = -\frac{GM_{BH}}{c^2r^3_{\star}}\biggl\{{\pmb{r}}_{\star}\biggl[c^2+v^2_{\star}+2v^2_{BH}-4\left({\pmb{v}}_{\star}.{\pmb{v}}_{BH}\right) \hspace{2.2cm}\\
-\frac{3}{2r^2_{\star}}\left({\pmb{r}}_{\star}.{\pmb{v}}_{BH}\right)^2-4\frac{GM_{BH}}{r_{\star}}\biggr] \hspace{2.2cm}\\
-\left[{\pmb{r}}_{\star}.\left(4{\pmb{v}}_{\star}-3{\pmb{v}}_{BH}\right) \right]\left({\pmb{v}}_{\star}-{\pmb{v}}_{BH}\right)\biggr\}. \hspace{1.6cm}
\end{split}
\end{equation}

\indent $M_{BH}$ is the dominant mass of the BH, ${\pmb{v}}_{\star}$ and ${\pmb{r}}_{\star}$ are the velocity and the radius vectors of the star, and ${\pmb{v}}_{\bullet}$ is the velocity vector of the BH. Here we consider only the mass of the BH since we are well inside the sphere of influence of the SMBH and hence we assume that the extended mass is negligible in comparison with the mass of the BH. Reducing the equation considering a negligible proper motion for the central BH gives us the equation of motion in \citet{rubilar2001}, which is given by

\begin{equation}
\label{eqn:geocm}   
\frac{d{\pmb{v}}_{\star}}{dt}=-\frac{GM_{BH}}{c^2r^3_{\star}}\biggl[{\pmb{r}}_{\star}\left(c^2-4\frac{GM_{BH}}{r_{\star}}+v^2_{\star}\right)
-4{\pmb{v}}_{\star}\left({\pmb{v}}_{\star}.{\pmb{r}}_{\star}\right)\biggr] \, ,  
\end{equation}

\noindent which can be used in the case where we are neglecting the small drift motion of the BH. We modeled the stellar orbits in Sect.~\ref{sec:orbits} or Sect.~\ref{sec:cases} by integrating Equations~\eqref{eqn:eih} or \eqref{eqn:geocm} using the fourth-order Runge--Kutta method with twelve or six initial parameters respectively (i.e. the positions and velocities in three dimensions).

\section{Stellar Orbits}
\label{sec:orbits}

\subsection{Astrometric Accuracy}
\label{sec:astro}
 
\indent \cite{gill2009} show that for raw positions and linear transformations, the resulting mean one-dimensional position error is as large as 1 mas for the S13 NACO data.

\indent \cite{plewa2015} find from the average velocity differences in radial and tangential directions that the infrared reference frame shows neither pumping (v$_r$/r) nor rotation ($v_{\phi}/r$) relative to the radio system to within $\sim$7.0~$\mu$as~yr$^{-1}$~arcsec$^{-1}$. Over 20~yr this amounts to an upper limit of about 0.14~mas~arcsec$^{-1}$, i.e. typically to 0.1--0.2~mas across the central 1~arcsec diameter cluster of high-velocity stars. This means that the combined error due to the residual distortions, the rotation, and the transformation across the central S-cluster is less than about 1.2~mas.

\indent The accuracy with which an individual stellar position can be derived via a Gaussian fit is better than a tenth of a pixel and ranges for the bright S-cluster stars between 1 and 2~mas per single epoch. Obtaining the position of Sgr~A* is complicated because of the crowding in the field and in particular due to the presence of S17 over a few years of our epochs. Hence, the accuracy in deriving the position of Sg~ A* typically ranges from 1 to 2~mas for the bright flare events and up to about 6 mas (about 1/2~pixel in camera S13) for fainter flare emissions and in the presence of S17.
\cite{plewa2015} have shown that accuracies of a fraction of a mas can be achieved  for sufficiently bright stars (see below).

\subsection{Connection of the NIR and Radio Reference Frames}
\label{sec:frames}

All instrumental imaging parameters
up to second order are extracted for each individual data
set. Here we assumed that the pixel coordinates of the
$i$th star ($x_i$, $y_i$) can be written in terms of the corrected offset
coordinates ($\Delta x_i$, $\Delta y_i$) from the base position as

\begin{equation}
x_i = a_0 + 
a_1 \Delta x_i  +
a_2 \Delta y_i  +
a_3 \Delta x_i^2  +
a_4 \Delta x_i \Delta y_i  +
a_5 \Delta y_i^2 
\end{equation}
and
\begin{equation}
y_i = b_0 + 
b_1 \Delta x_i  +
b_2 \Delta y_i  +
b_3 \Delta x_i^2  +
b_4 \Delta x_i \Delta y_i  +
b_5 \Delta y_i^2~~~. 
\end{equation}

\indent The zeroth order is the base position ($a_0$, $b_0$), the first order
(proportional to $\Delta x$ and $\Delta y$ and in each coordinate) relates to
the camera rotation angle $\alpha_r$ and the pixel scales 
$\rho_x$, $\rho_y$ (in arcsec pixel$^{-1}$), and the second-order parameters (proportional to $\Delta x^2$, $\Delta x$ $\Delta y$, and $\Delta y^2$ for each coordinate) give the
image distortions. The 2~$\times$~6 instrumental parameters 
($a_0$, b$_0$, ..., $a_5$, $b_5$) are determined for each data set by comparison
to a radio reference frame consisting of eight maser sources
with positions and proper motions tabulated by \cite{plewa2015}.
The parameters are computed by solving an
over-determined nonlinear equation for these eight stars
via the orthonormalization of the 12 $\times$~8 matrix. Based on this
analysis we find that the fitted pixel scales and the very
small second-order distortion parameters are typically
$<$~10$^{-3}$ of the first-order scaling parameters. 
After the correction for the instrumental
parameters the final errors of the position fit ranged from 1 to 10 mas per data set for the bright maser stars.

\indent For each year we choose a wide-field (27~mas~pix$^{-1}$ scale) image containing the maser sources 
and the closest high-resolution (13~mas~pix$^{-1}$ scale) image in which the S-stars and 
Sgr~A* can best be separated. Via the formalism given above we then match the higher~resolution 13~mas~
pixel$^{-1}$ scale positions onto the larger-field 27~mas~pix$^{-1}$ scale images containing the infrared
counterparts of the maser sources.
The corresponding frames are connected using six overlap sources
for which the offsets to Sgr~A* are measured:
IRS16SW (S95), IRS16C (S97), S65, S96, S67, and S2. In a second step we use the distortion-corrected infrared positions 
(i.e. their projected trajectories) of the radio maser star
counterparts given by \cite{plewa2015} to connect our positional reference frame to the radio frame.
This is done under the assumption that 
the radio masers are quasi-co-spatial with the associated stars.
\cite{oyama2008} and \citet{sjouwerman2004}
show that this is a reasonable assumption because the maser spot shells are distributed over 
less than 1 mas around their central stars.

\indent As a result we find the motion of the infrared counterpart of Sgr~A* with respect to the radio rest frame.
We find that (over our baseline in time) the infrared position of Sgr~A* agrees with the 
radio position to within less than 1.4~mas
and the proper motion is smaller than 0.3~mas per year.
Hence, this is the accuracy with which we can connect the infrared and radio
reference frames for the central S-star cluster, which is about an order of magnitude 
below what has been achieved by \cite{plewa2015}, i.e., $\sim$~0.17 mas 
in position (in 2009) and $\sim$0.07~mas~$yr^{-1}$ in velocity. 
Hence, the comparison to the radio reference frame shows that 
the infrared and radio positions of Sgr~A* are in good agreement and that to the first order the S-stars are orbiting the IR
counterpart of Sgr~A*.

\begin{figure}[htbp]
\centering
\subfigure{\includegraphics[width=0.5\textwidth]{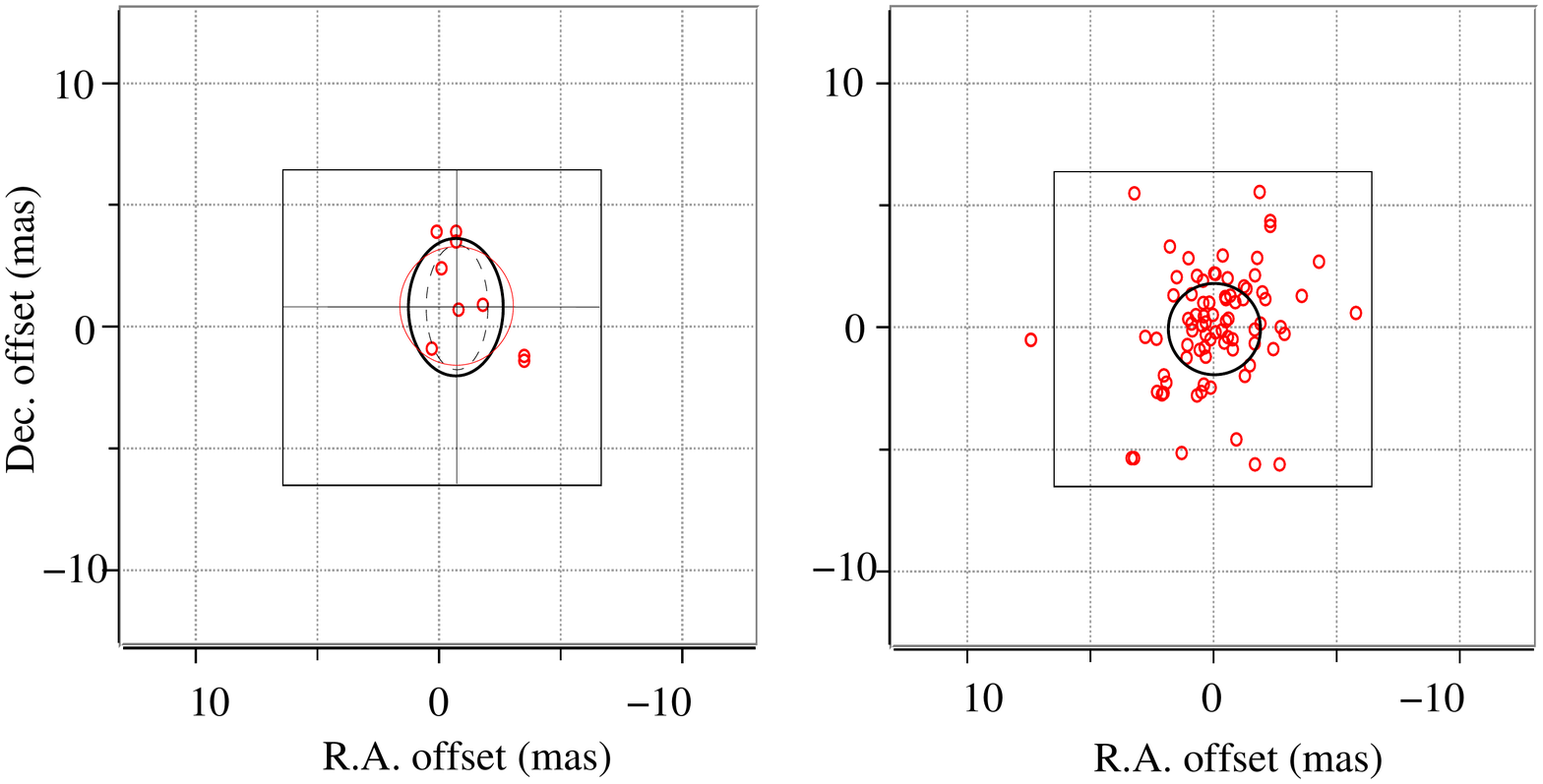}}
\caption{Left: Single-epoch statistics for the offset between the infrared
and radio positions of Sgr~A*.
The uncertainties for the R.A. and Dec.:
With respect to the median offset the zero-offset point is well included in
the median deviation: 1.8~mas~$\times$~0.9~mas (thin red ellipse);
the standard deviation: 2.0~mas~$\times$~1.4~mas (thick black ellipse);
the equivalent geometrical mean: 1.7~mas (black dashed ellipse).
Right: Single-epoch statistics for all maser sources well centered on the
 zero-offset point. The standard deviation is 1.8~mas (black circle).}
\label{fig:newsgr1}
\end{figure}

\indent We can compare our result with the expectations from the input data.
If we consider that, depending on the source strength, the stellar positions have been measured from the NIR images
with an accuracy of better than 1--10~mas (typically better than between 0.037 and 0.3~pixels) then the uncertainty in the connection 
to the radio frame is dominated by the correction for the distortion of
about 1~mas, as corrected by \cite{plewa2015}.
In the following we will stay with the conservative assumption of an accuracy of 10~mas for the 
position determination.
The remaining uncertainty in the connection 
to the radio frame is influenced by: 1) The accuracy in mosaicking to access the maser source in the 27~mas~pix$^{-1}$ fields; this process is typically
affected by of the order of nine sources along the overlap regions between the frames;
2) measuring from the eight maser sources in the 27~mas~pix$^{-1}$ mosaics; 3) connecting the 27~mas~pix$^{-1}$ scale to the 13~mas~pix$^{-1}$ scale data using six sources.
We assume that the accuracy in determining positions in the 13~mas~pix$^{-1}$ scale field is twice as large as in the 27~mas~pix$^{-1}$ scale field.
As a result the final accuracy for the determination of a single source position is of the order of half the
accuracy reached in the single 27~mas~pix$^{-1}$ scale frames, i.e., between 0.5 and~5 mas. Using all eight maser sources to determine
the single~epoch positions for the measurements of proper motion gives an accuracy of at least 1.7~mas. As shown in Fig.~\ref{fig:newsgr1} the single-epoch statistics for Sgr~A* and for
all eight maser sources are in very good agreement with this estimate.
Our analysis of the S-star orbits below shows that 
we achieved a positional accuracy on the comparatively faint S-cluster sources of 3~mas.

\subsection{Derivation of the Positions and the Orbits}
\label{sec:orbitsderiv}

\indent The procedure described above allows us to derive the stellar positions and the infrared position of Sgr~A* with respect to the radio rest frame. We choose only images (33 in total) in which Sgr~A* was flaring in the infrared in order to locate it directly in our coordinate system and have a good control of its possible motion with respect to the stellar cluster.
In addition to the three candidate stars (S2, S38, and S55/S0-102), five stars (S7, S10, S26, S30, and S65) in their vicinity of them are selected to verify their reported positions and motions \citep{gill2009, plewa2015, gill17}.
This allowed us to validate the quality of the reference frame on an image-by-image basis.
These reference stars are chosen from the bright sources within the central arcsecond and have relatively low velocities and almost linear motions on the sky. Moreover, they can always be detected without any confusion or overlapping with other sources.

\indent After locating all sources in all images, the pixel positions of the three candidate stars and five reference stars are extracted. This is done by two-dimensional Gaussian fits of the position of the isolated sources. In the case of partial overlapping of stars in some epochs, the pixel positions are obtained without Gaussian fit, and therefore larger corresponding measurement errors are considered.

\indent S2 can be detected in all 33 images from 2002 to 2015. S38 is probably confused with other sources in the years before 2004 and thus we keep only 29 astrometric measurements from 2004 to 2015 for it. Also S55/S0-102 is a faint star \citep[16 times fainter than S2;][]{meyer12} located in a very crowded region close to Sgr~A*, therefore it is not detectable in every image, and that leaves us with 25 measurements from 2004 to 2015.

\indent The pixel positions are transformed into an astronomic reference frame. To do so, we fit a linear equation of motion to the five reference stars, given in Table \ref{table:five}, and find the residuals in all the mosaics. We use the mean of all the residuals from all the reference stars in the image to correct for the image distortion and the astronomic positions of all our sources including Sgr~A* in each corresponding image. The standard deviation of the mean of the residuals is inserted into the uncertainties of the astrometric data as the uncertainty of the reference frame. The resulting positions are given in Tables~\ref{table:s2pos}--\ref{table:s0102pos} for S2, S38, and S0-102/S55, respectively. Fitting a linear motion to the Sgr~A* data after these transformations, we get

\begin{equation}
\label{eqn:sgralinemotion}
\begin{split}
\alpha(mas) & = (1.9 \pm 2.5) - (0.21 \pm 0.37)\times(t(yr) - 2002.578) \\
\delta(mas) & = (-0.4 \pm 2.4) + (0.06 \pm 0.41)\times(t(yr) - 2002.578).
\end{split}
\end{equation}

\begin{table*}[htbp]
\centering
\caption{Equation of motion of the five reference stars.}
\label{table:five}
\begin{tabular}{l c c c}
\hline\hline
Star & $t_0$ & $\Delta$R.A. & $\Delta$Dec. \\
     & (yr)  & (arcsec)     & (arcsec)     \\
\hline
S7  & 2002.578 & (0.5146 $\pm$ 0.0026)+(-0.0040 $\pm$ 0.0001)$\Delta$t & (-0.0421 $\pm$ 0.0020)+(-0.0016 $\pm$ 0.0002)$\Delta$t \\
S10 & 2002.578 & (0.0552 $\pm$ 0.0023)+(-0.0045 $\pm$ 0.0001)$\Delta$t & (-0.3736 $\pm$ 0.0020)+(0.0037 $\pm$ 0.0002)$\Delta$t \\
S26 & 2002.578 & (0.5105 $\pm$ 0.0027)+(0.0060 $\pm$ 0.0001)$\Delta$t & (0.4296 $\pm$ 0.0020)+(0.0016 $\pm$ 0.0004)$\Delta$t \\
S30 & 2002.578 & (-0.5434 $\pm$ 0.0024)+(-0.0001 $\pm$ 0.0003)$\Delta$t & (0.3806 $\pm$ 0.0021)+(0.0036 $\pm$ 0.0002)$\Delta$t \\
S65 & 2002.578 & (-0.7575 $\pm$ 0.0034)+(0.0023 $\pm$ 0.0006)$\Delta$t & (-0.2684 $\pm$ 0.0033)+(-0.0015 $\pm$ 0.0006)$\Delta$t \\
\hline
\end{tabular}
\end{table*}

\begin{table}[htbp]
\centering
\caption{Astrometric Measurements of S2}
\label{table:s2pos}
\begin{tabular}{l c c c c}
\hline\hline
Date & $\Delta$R.A. & $\Delta$Dec. & $\Delta$R.A. Error & $\Delta$Dec. Error \\
(Decimal) & (arcsec) & (arcsec) & (arcsec) & (arcsec) \\
\hline
2002.578 & 0.0386 & 0.0213 & 0.0066 & 0.0065 \\
2003.447 & 0.0385 & 0.0701 & 0.0009 & 0.0010 \\
2003.455 & 0.0393 & 0.0733 & 0.0012 & 0.0012 \\
2004.511 & 0.0330 & 0.1191 & 0.0010 & 0.0008 \\
2004.516 & 0.0333 & 0.1206 & 0.0009 & 0.0006 \\
2004.574 & 0.0315 & 0.1206 & 0.0009 & 0.0009 \\
2005.268 & 0.0265 & 0.1389 & 0.0007 & 0.0011 \\
2006.490 & 0.0141 & 0.1596 & 0.0065 & 0.0065 \\
2006.584 & 0.0137 & 0.1609 & 0.0033 & 0.0007 \\
2006.726 & 0.0129 & 0.1627 & 0.0033 & 0.0007 \\
2006.800 & 0.0107 & 0.1633 & 0.0033 & 0.0007 \\
2007.205 & 0.0064 & 0.1681 & 0.0004 & 0.0007 \\
2007.214 & 0.0058 & 0.1682 & 0.0004 & 0.0008 \\
2007.255 & 0.0069 & 0.1691 & 0.0010 & 0.0007 \\
2007.455 & 0.0047 & 0.1709 & 0.0004 & 0.0006 \\
2008.145 & -0.0076 & 0.1775 & 0.0007 & 0.0012 \\
2008.197 & -0.0082 & 0.1780 & 0.0007 & 0.0011 \\
2008.268 & -0.0084 & 0.1777 & 0.0006 & 0.0008 \\
2008.456 & -0.0118 & 0.1798 & 0.0006 & 0.0009 \\
2008.598 & -0.0126 & 0.1802 & 0.0009 & 0.0010 \\
2008.708 & -0.0127 & 0.1806 & 0.0008 & 0.0013 \\
2009.299 & -0.0216 & 0.1805 & 0.0006 & 0.0009 \\
2009.334 & -0.0218 & 0.1813 & 0.0006 & 0.0009 \\
2009.501 & -0.0233 & 0.1803 & 0.0005 & 0.0008 \\
2009.605 & -0.0266 & 0.1800 & 0.0012 & 0.0015 \\
2009.611 & -0.0249 & 0.1806 & 0.0006 & 0.0008 \\
2009.715 & -0.0260 & 0.1804 & 0.0006 & 0.0008 \\
2010.444 & -0.0347 & 0.1780 & 0.0013 & 0.0021 \\
2010.455 & -0.0340 & 0.1774 & 0.0008 & 0.0013 \\
2011.400 & -0.0430 & 0.1703 & 0.0009 & 0.0017 \\
2012.374 & -0.0518 & 0.1617 & 0.0012 & 0.0016 \\
2013.488 & -0.0603 & 0.1442 & 0.0006 & 0.0019 \\
2015.581 & -0.0690 & 0.1010 & 0.0014 & 0.0010 \\
\hline
\end{tabular}
\end{table}

\begin{table}[htbp]
\centering
\caption{Astrometric Measurements of S38}
\label{table:s38pos}
\begin{tabular}{l c c c c}
\hline\hline
Date & $\Delta$R.A. & $\Delta$Dec. & $\Delta$R.A. Error & $\Delta$Dec. Error \\
(Decimal) & (arcsec) & (arcsec) & (arcsec) & (arcsec) \\
\hline
2004.511 & -0.0667 & 0.0576 & 0.0017 & 0.0016 \\
2004.516 & -0.0673 & 0.0690 & 0.0066 & 0.0065 \\
2005.268 & -0.1178 & 0.0583 & 0.0065 & 0.0066 \\
2006.490 & -0.1544 & 0.0558 & 0.0065 & 0.0065 \\
2006.584 & -0.1600 & 0.0613 & 0.0073 & 0.0078 \\
2006.726 & -0.1684 & 0.0550 & 0.0009 & 0.0008 \\
2006.800 & -0.1690 & 0.0549 & 0.0011 & 0.0009 \\
2007.205 & -0.1851 & 0.0513 & 0.0005 & 0.0008 \\
2007.214 & -0.1853 & 0.0506 & 0.0005 & 0.0008 \\
2007.255 & -0.1807 & 0.0524 & 0.0010 & 0.0007 \\
2007.455 & -0.1898 & 0.0474 & 0.0005 & 0.0065 \\
2008.145 & -0.2058 & 0.0363 & 0.0009 & 0.0013 \\
2008.197 & -0.2065 & 0.0359 & 0.0008 & 0.0011 \\
2008.268 & -0.2049 & 0.0338 & 0.0007 & 0.0009 \\
2008.456 & -0.2111 & 0.0325 & 0.0008 & 0.0010 \\
2008.598 & -0.2141 & 0.0346 & 0.0010 & 0.0010 \\
2008.708 & -0.2175 & 0.0338 & 0.0010 & 0.0013 \\
2009.299 & -0.2315 & 0.0244 & 0.0007 & 0.0009 \\
2009.334 & -0.2310 & 0.0241 & 0.0007 & 0.0009 \\
2009.501 & -0.2344 & 0.0216 & 0.0006 & 0.0008 \\
2009.605 & -0.2360 & 0.0156 & 0.0012 & 0.0015 \\
2009.611 & -0.2350 & 0.0202 & 0.0007 & 0.0008 \\
2009.715 & -0.2363 & 0.0178 & 0.0006 & 0.0009 \\
2010.444 & -0.2415 & 0.0053 & 0.0013 & 0.0021 \\
2010.455 & -0.2437 & 0.0009 & 0.0009 & 0.0014 \\
2011.400 & -0.2425 & -0.0113 & 0.0010 & 0.0017 \\
2012.374 & -0.2519 & -0.0251 & 0.0013 & 0.0017 \\
2013.488 & -0.2450 & -0.0409 & 0.0007 & 0.0019 \\
2015.581 & -0.2320 & -0.0617 & 0.0016 & 0.0013 \\
\hline
\end{tabular}
\end{table}

\begin{table}[htbp]
\centering
\caption{Astrometric MeasurementsS0-102/S55}
\label{table:s0102pos}
\begin{tabular}{l c c c c}
\hline\hline
Date & $\Delta$R.A. & $\Delta$Dec. & $\Delta$R.A. Error & $\Delta$Dec. Error \\
(Decimal) & (arcsec) & (arcsec) & (arcsec) & (arcsec) \\
\hline
2004.511 & 0.0549 & -0.1552 & 0.0066 & 0.0065 \\
2004.516 & 0.0711 & -0.1536 & 0.0066 & 0.0065 \\
2005.268 & 0.0707 & -0.1437 & 0.0065 & 0.0066 \\
2006.490 & 0.0731 & -0.1199 & 0.0065 & 0.0065 \\
2006.584 & 0.0749 & -0.1220 & 0.0065 & 0.0065 \\
2006.726 & 0.0790 & -0.1180 & 0.0066 & 0.0065 \\
2006.800 & 0.0731 & -0.1169 & 0.0066 & 0.0065 \\
2007.205 & 0.0835 & -0.0883 & 0.0065 & 0.0065 \\
2007.255 & 0.0797 & -0.0763 & 0.0066 & 0.0065 \\
2007.455 & 0.0784 & -0.0635 & 0.0065 & 0.0065 \\
2008.145 & 0.0659 & -0.0346 & 0.0065 & 0.0066 \\
2008.197 & 0.0641 & -0.0338 & 0.0065 & 0.0066 \\
2008.268 & 0.0711 & -0.0309 & 0.0065 & 0.0066 \\
2008.456 & 0.0692 & -0.0167 & 0.0065 & 0.0066 \\
2008.598 & 0.0678 & -0.0144 & 0.0066 & 0.0066 \\
2008.708 & 0.0620 & -0.0058 & 0.0066 & 0.0066 \\
2009.334 & -0.0017 & 0.0358 & 0.0065 & 0.0066 \\
2009.501 & -0.0257 & 0.0291 & 0.0065 & 0.0066 \\
2009.605 & -0.0305 & 0.0243 & 0.0066 & 0.0067 \\
2009.715 & -0.0390 & 0.0378 & 0.0065 & 0.0066 \\
2010.444 & -0.0620 & -0.0453 & 0.0066 & 0.0068 \\
2010.455 & -0.0523 & -0.0404 & 0.0018 & 0.0020 \\
2011.400 & -0.0492 & -0.1080 & 0.0066 & 0.0067 \\
2012.374 & -0.0345 & -0.1180 & 0.0013 & 0.0029 \\
2013.488 & -0.0134 & -0.1380 & 0.0007 & 0.0019 \\
2015.581 & 0.0239 & -0.1678 & 0.0016 & 0.0066 \\
\hline
\end{tabular}
\end{table}

\begin{figure}[htbp]
\centering
\subfigure{\includegraphics[width=0.4\textwidth]{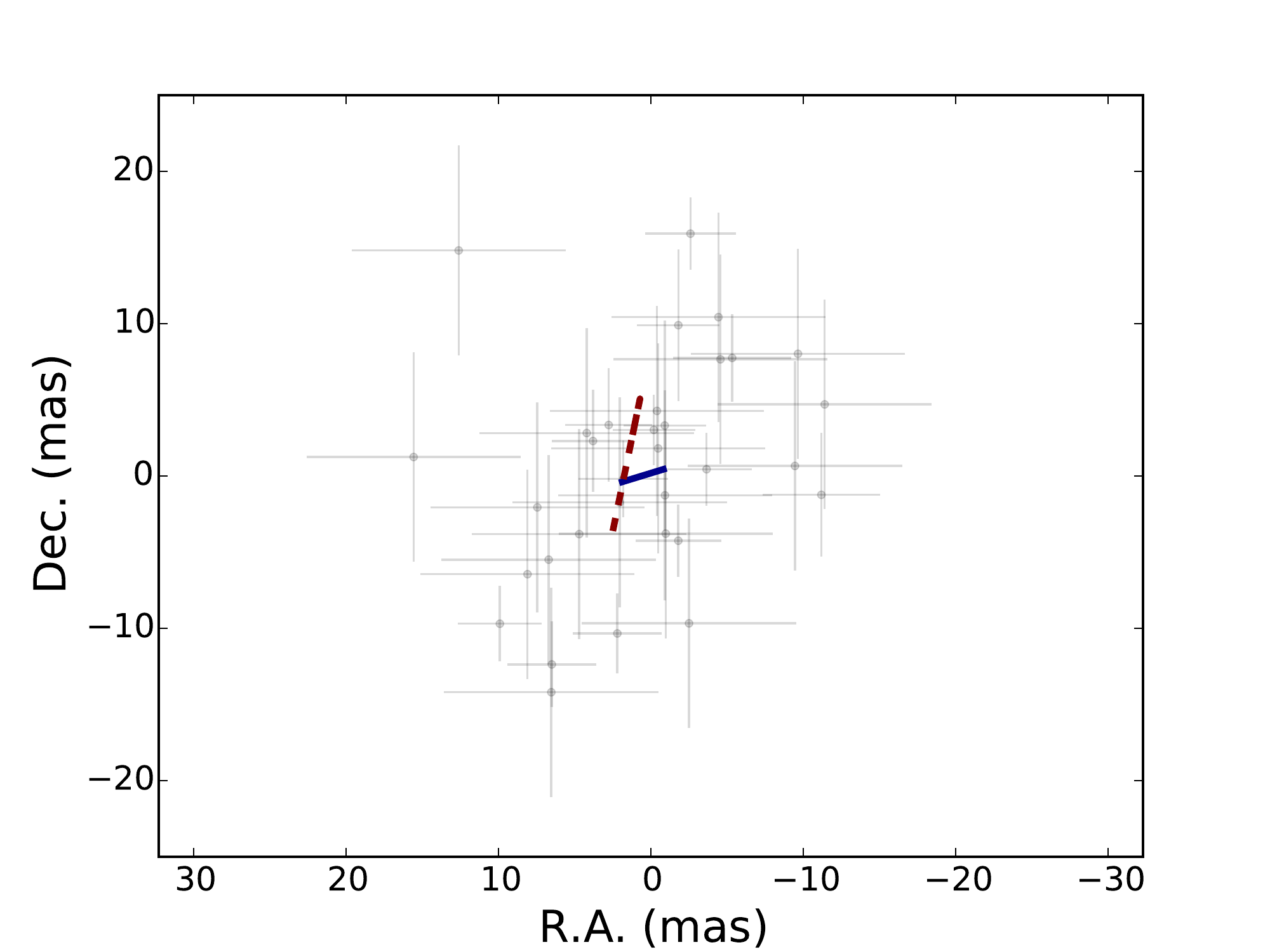}}
\caption{Linear motion fit to the  data on the NIR counterpart of Sg~A*(derived from Newtonian orbit fitting to all three stars) after applying the correction described in the text in this study (solid blue) compared to a recent study (dashed red, \citet{gill2009}). The data points with crosses indicating their uncertainties are the positions we derived for the IR counterpart of Sgr~A*.}
\label{fig:sgra}
\end{figure}

Figure \ref{fig:sgra} shows this linear fit compared to the previous study done by \citet{gill2009}. The linear fits and the uncertainties of the fits were done with a bootstrap algorithm in which we generate 50 random samples with replacements of size equal to the observed dataset and calculate the statistics on the sampling distribution.

\begin{figure*}[htbp]
\centering
\subfigure{\includegraphics[width=1\textwidth]{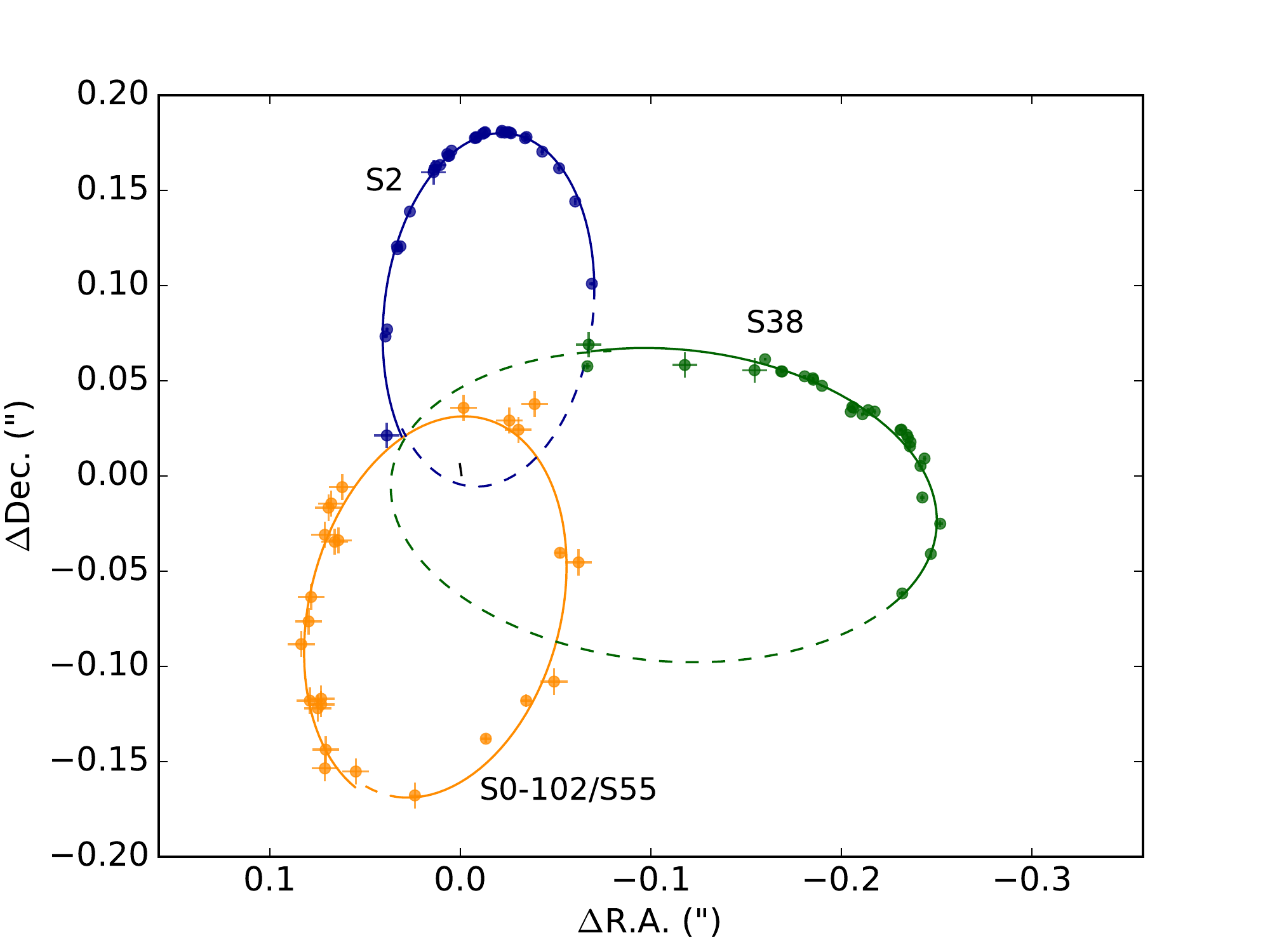}}
\caption{Best relativistic fit orbit of the candidate stars using equation \eqref{eqn:eih} and the minimized ${\chi}^2$ method. The astrometric data in the reference coordinate system are represented by points with error bars (smaller than diameter of the point in most cases). The relativistic orbits of the fits are shown by solid lines. Extrapolations before and after the 
region for which we have data are shown as dashed lines. The data (from 2002 to 2015) and the orbit of S2 are represented in blue, the S55/S0-102 data (from 2004 to 2015) and its orbit are orange, the S38 data (from 2004 to 2015) and its orbit are green. The motion of Sgr A* as derived from the relativistic fits to all three stars is shown in black.}
\label{fig:all}
\end{figure*}

\indent We use the PN approximation discussed in Sect.~\ref{sec:simul} and fit our astrometric data for the candidate stars (and simultaneously for the radial velocity in the case of S2 and S38) to the relativistic model using the minimum ${\chi}^2$ method. The measurement errors (considering both astrometric errors and reference frame errors) are scaled in a such way that the reduced $\chi^2$~=~1. The result is shown in Fig.~\ref{fig:all} for all candidate stars.

\begin{figure*}[htbp]
\centering
\subfigure{\includegraphics[width=1\textwidth]{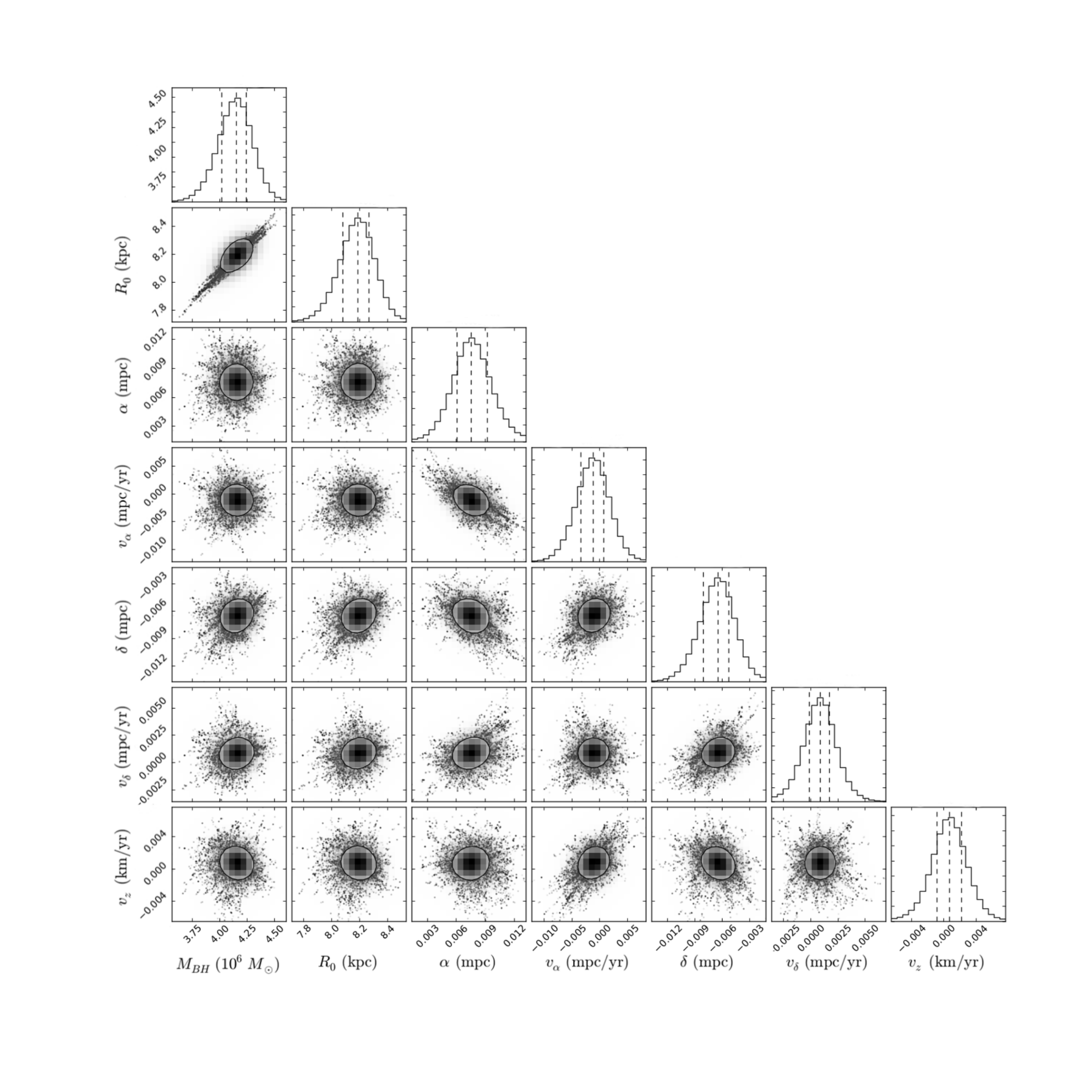}}
\caption{Gravitational potential parameters from the simultaneous fit to all the candidate stars (S2, S38, and S55/S0-102) and the gravitational potential parameters including the mass and the distance to the SMBH using MCMC simulations and Newtonian models for the stars. The remaining parameters are not shown so as to allow for readability of the displayed graphs. Each panel shows a 2D cut of the parameter space. The posterior probability distribution is compact. The marginalized distribution for each parameter is shown independently in the histograms along the diagonal. The contours show the 1$\sigma$ uncertainties in the 2D histograms and the dashed lines show the 0.16, 0.5, and 0.84 quantiles in the 1D histograms.}
\label{fig:mcmcall}
\end{figure*}

\indent Moreover, we run Markov Chain Monte Carlo (MCMC) simulations using Newtonian models to find all six orbital parameters for the candidate stars and the gravitational potential parameters of the SMBH (including its mass and the distance to the GC), and their 1$\sigma$ uncertainties simultaneously. We use \textit{emcee} by \citet{emcee}, which is an affine-invariant ensemble sampler for MCMC. We repeat the simulations using one (S2), two (S2 and S38), and three (S2, S38, and S55/S0-102) stars. Figure~\ref{fig:mcmcall} is a part of the results for such a simulation. The rest of the parameters are omitted for reasons of legibility. The histograms along the diagonal are the marginalized distribution for each parameter and resemble normal distributions. The rest of the panels show 2D cuts of the parameter space. If the posterior probability is broad then that parameter is not well constrained. However, the posterior probability is compac,t which means all parameters are well constrained. There are some correlations between the parameters, especially between $M_{BH}$ and $R_0$. 

\indent We change the model afterwards to a relativistic one using Equation \eqref{eqn:eih} and repeat the simulations for the same combinations of the candidate stars. The results from all MCMC simulations are given in Table \ref{table:allresults}. The errors of the parameters are the result of their distributions from the MCMC simulations. Using two and three candidate stars reduces the uncertainties in determining most of the parameters, especially the velocity of Sgr~A*, as a result of the lack of astrometric data in the lower part of the orbit of S2 and S38. Moreover \citet{zuc2006} indicate that using a Keplerian model instead of a relativistic one might systematically underestimate $R_0$.

\begin{table*}[htbp]
\centering
\caption{Results of the MCMC simulations considering both Keplerian and relativistic models for the stars. Different combination of S2, S38 and S55/S0-102 were used for both models. Using two and three stars instead of only S2 reduces the uncertainties of all parameters. Due to the facts that S38 has been observed for only half an orbit and the lack of data on the radial velocity on S55/S0-102, other combinations of these three stars lead to poorly determined parameters, therefore they are not included in this table. 
The reference epochs for the relativistic and Keplerian fits are 2002 April and 2002 July, respectively. This needs to be considered when comparing the positions of the SMBH.}
\label{table:allresults}
\begin{tabular}{l r r r r r r}
\hline\hline
                        &                             & Keplerian                    &                     &         & Relativistic &                   \\
Parameter (unit)        & S2 Only                     & S2, S38                   & \textbf{S2, S38, S55/S0-102} & S2 Only & S2, S38   & S2, S38, S55/S0-102\\
\hline
Black hole:\\
$M_{BH}$~(10$^6$ \msun) & $4.78^{+0.73}_{-0.47}$   & $4.08^{+0.17}_{-0.17}$    & $\mathbf{4.15^{+0.09}_{-0.13}}$ & $4.37^{+0.20}_{-0.14}$ & $4.16^{+0.02}_{-0.02}$& $4.72^{+0.08}_{-0.06}$ \\
Distance (kpc)          & $8.93^{+0.57}_{-0.44}$   & $8.14^{+0.13}_{-0.12}$    & $\mathbf{8.19^{+0.08}_{-0.11}}$ & $8.50^{+0.16}_{-0.18}$ & $8.29^{+0.01}_{-0.01}$& $8.53^{+0.07}_{-0.03}$ \\
$\alpha$ (mas)          & $0.31^{+0.95}_{-0.73}$   & $0.21^{+0.04}_{-0.05}$    & $\mathbf{0.19^{+0.04}_{-0.04}}$ & $-0.03^{+0.17}_{-0.14}$&$0.03^{+0.07}_{-0.14}$& $0.07^{+0.08}_{-0.03}$\\
$\delta$ (mas)          & $0.42^{+1.32}_{-0.82}$   & $0.23^{+0.10}_{-0.10}$    & $\mathbf{-0.16^{+0.03}_{-0.41}}$ & $-0.01^{+0.06}_{-0.07}$&$-0.40^{+0.04}_{-0.05}$& $0.56^{+0.10}_{-0.08}$ \\
$v_\alpha$ (mas~yr$^{-1}$)     & $-0.21^{+0.36}_{-0.42}$  & $-0.11^{+0.15}_{-0.18}$   & $\mathbf{-0.03^{+0.05}_{-0.06}}$& $-0.07^{+0.11}_{-0.16}$& $0.56^{+0.05}_{-0.03}$& $0.19^{+0.12}_{-0.06}$ \\
$v_\delta$ (mas~yr$^{-1}$)     & $0.13^{+1.02}_{-0.66}$   & $0.06^{+0.17}_{-0.13}$    & $\mathbf{0.02^{+0.02}_{-0.03}} $& $0.12^{+0.23}_{-0.15}$ & $-0.08^{+0.08}_{-0.14}$& $0.34^{+0.05}_{-0.04}$ \\
$v_z$      (km~s$^{-1}$)       &$-10.3^{+50.7}_{-43.0}$& $-2.01^{+5.97}_{-7.84}$   & $\mathbf{0.70^{+1.47}_{-1.52}}$ &$-3.17^{+3.46}_{-2.05}$&$22.30^{+1.42}_{-2.08}$& $18.81^{+4.78}_{-9.00}$\\
S2:\\
$a$ ($\arcsec$)                   & $0.121^{+0.006}_{-0.004}$ & $0.126^{+0.002}_{-0.002}$ & $\mathbf{0.126^{+0.001}_{-0.001}}$\\
$e$                       & $0.872^{+0.006}_{-0.007}$ & $0.882^{+0.003}_{-0.004}$ & $\mathbf{0.884^{+0.002}_{-0.002}}$\\
$i$ ($\degr$)         & $138.1^{+2.0}_{-1.8}$ & $136.38^{+0.77}_{-0.91}$ & $\mathbf{136.78^{+0.36}_{-0.44}} $\\
$\omega$ ($\degr$)  & $68.9^{+1.9}_{-1.9}$ & $71.1^{+1.3}_{-1.4}$ & $\mathbf{71.36^{+0.65}_{-0.84}}$\\
$\Omega$ ($\degr$)  & $231.9^{+2.8}_{-2.6}$ & $233.9^{+1.7}_{-1.9}$ & $\mathbf{234.50^{+0.94}_{-1.09}}$\\
$T_p$ (yr)              & $2002.27^{+0.04}_{-0.04}$& $2002.33^{+0.02}_{-0.02}$ & $\mathbf{2002.32^{+0.02}_{-0.02}}$\\
S38:\\
$a$ ($\arcsec$)                   &                             & $0.139^{+0.002}_{-0.003}$    & $\mathbf{0.140^{+0.001}_{-0.002}}$&   \\
$e$                       &                             & $0.819^{+0.005}_{-0.005}$    & $\mathbf{0.818^{+0.005}_{-0.005}}$ &  \\
$i$ ($\degr$)         &                             & $167.1^{+2.6}_{-2.6}$  & $\mathbf{166.22^{+3.1}_{-2.4}}$& \\
$\omega$ ($\degr$)  &                             & $27.5^{+9.8}_{-7.4}$   & $\mathbf{18.4^{+4.8}_{-5.8}}$&  \\
$\Omega$ ($\degr$)  &                             & $106.8^{+9.5}_{-7.2}$  & $\mathbf{101.8^{+4.6}_{-5.6}}$& \\
$T_p$ (yr)              &                             & $2003.32^{+0.03}_{-0.04}$ & $\mathbf{2003.30^{+0.03}_{-0.04}}$&\\
S55/S0-102:\\
$a$ ($\arcsec$)                   &                             &                              & $\mathbf{0.109^{+0.002}_{-0.002}}$   \\
$e$                       &                             &                              & $\mathbf{0.74^{+0.01}_{-0.01}}$   \\
$i$ ($\degr$)         &                             &                              & $\mathbf{141.7^{+1.6}_{-1.5}}$ \\
$\omega$ ($\degr$)  &                             &                              & $\mathbf{133.5^{+3.9}_{-3.6}}$ \\
$\Omega$ ($\degr$)  &                             &                              & $\mathbf{129.9^{+4.0}_{-4.2}}$ \\
$T_p$ (yr)              &                             &                              & $\mathbf{2009.31^{+0.03}_{-0.03}}$\\
\hline
\end{tabular}
\end{table*}

\indent For the starting point of the MCMC simulations, first we minimize the ${\chi}^2$ for the orbital parameters (the position and velocity for the relativistic model at the reference epoch) and only the mass of the SMBH and the distance to the GC. Then we try to improve the resulting ${\chi}^2$ by setting the position and the velocity of the central mass as free parameters. The results are then used in the code for MCMC simulations. The reference epoch for the relativistic fits is 2002 April and that for the Newtonian fits is 2002 July. This needs to be considered when comparing the position of the BH at the reference epochs of the two models. 

\section{Simulations and Case Studies}
\label{sec:cases}

\subsection{The Case of Simulated Stars within the Orbit of S2}
\label{sec:sim}

\indent The orbits of 14 stars (see Table \ref{table:noi}) are generated using Equation \eqref{eqn:geocm} by positioning them at different apoapse distances with different velocities within the orbit of S2 and integrating the equation of motion using the fourth-order Runge--Kutta method until the next apoapse is reached. Since the eccentricity of the orbit is not one of the initial parameters, an additional parameter $\alpha \equiv r \times v^{2}$ is introduced that has a linear correlation with the eccentricity in order to be able to generate orbits with the same eccentricity. If the total energy does not remain constant, the orbit gets stretched over time and the next apoapse will not be equal to the first one; consequently the resulting orbit will fail to be suitable for the purpose of this study. Hence, in order to keep the energy constant to a desirable approximation for the first few orbits, the time steps are taken to be relatively small. The drift motion of the BH can potentially have a large effect on the simulated orbits, but small orbits with short orbital periods have the advantage of being immune to the possible effects of the motion of the central mass.

\indent Also we assume that each S-star is a single star and not a binary or a bright component of one. Binaries can have effects on astrometric and radial velocity measurements and the binary disruption at the periapse can affect the orbit of the primary. Moreover there is no evidence, at least for S2, of a secondary component in the spectra \citep{eisen5, martins7, martins8}.

\begin{table*}[htbp]
\centering
\caption{Simulated case studies generated in section~\ref{sec:cases}. The second and third columns are the initial positions and velocities. The subscripts $u$ and $l$ are standing for 
the fit to the upper and lower parts of the orbit, respectively. The last column is the periapse shift. The last row shows the results for the case of S2 using the data from this study and the data published in \citet{boehle16} brought together in a same coordinate system. For all cases presented here, the uncertainties were calculated by the MCMC method.}
\label{table:noi}
\begin{tabular}{l c c c c c c c c c}
\hline\hline
Star & $\Delta$R.A.  & ($v_{Dec.}$ , $v_{z}$) & \ups & $a_u$ & $a_l$ & $e_u$ & $e_l$ & $\Delta \omega$ \\
     & (mpc)  & (10$^3$km~s$^{-1}$)  & & (mpc) & (mpc) &  &  & (rad) \\
\hline
 1  & 0.07 &  (5.34 , 0.00)    & 0.120 & $0.0252 \pm 0.0004$ & $0.0173 \pm 0.0017$ & $0.947 \pm 0.005$ & $0.823 \pm 0.024$ & $0.586 \pm 0.029$\\
 2  & 0.5  &  (2.00 , 0.00)    & 0.015 & $0.2573 \pm 0.0002$ & $0.2386 \pm 0.0039$ & $0.909 \pm 0.001$ & $0.885 \pm 0.003$ & $0.070 \pm 0.001$\\
 3  & 0.05 &  (7.75 , 0.00)    & 0.106 & $0.0210 \pm 0.0003$ & $0.0166 \pm 0.0012$ & $0.915 \pm 0.007$ & $0.789 \pm 0.014$ & $0.530 \pm 0.014$\\
 4  & 0.06 &  (7.07 , 0.00)    & 0.087 & $0.0269 \pm 0.0002$ & $0.0209 \pm 0.0013$ & $0.907 \pm 0.004$ & $0.788 \pm 0.012$ & $0.434 \pm 0.016$\\
 5  & 0.04 & (10.00 , 0.00)    & 0.095 & $0.0185 \pm 0.0002$ & $0.0154 \pm 0.0005$ & $0.871 \pm 0.004$ & $0.735 \pm 0.008$ & $0.480 \pm 0.012$\\
 6a & 0.06 &  (6.45 , 5.00)    & 0.061 & $0.0300 \pm 0.0006$ & $0.0260 \pm 0.0021$ & $0.840 \pm 0.005$ & $0.741 \pm 0.019$ & $0.301 \pm 0.023$\\
 6b & 0.06 &  (8.16 , 0.00)    & 0.061 & $0.0300 \pm 0.0002$ & $0.0259 \pm 0.0007$ & $0.850 \pm 0.002$ & $0.748 \pm 0.001$ & $0.312 \pm 0.008$\\
 7a & 0.05 &  (8.00 , 6.00)    & 0.056 & $0.0262 \pm 0.0003$ & $0.0232 \pm 0.0013$ & $0.814 \pm 0.008$ & $0.705 \pm 0.011$ & $0.279 \pm 0.020$\\
 7b & 0.05 & (10.00 , 0.00)    & 0.056 & $0.0262 \pm 0.0001$ & $0.0238 \pm 0.0005$ & $0.803 \pm 0.003$ & $0.702 \pm 0.005$ & $0.294 \pm 0.006$\\
 8  & 0.07 &  (8.45 , 0.00)    & 0.039 & $0.0379 \pm 0.0001$ & $0.0355 \pm 0.0008$ & $0.786 \pm 0.003$ & $0.712 \pm 0.008$ & $0.200 \pm 0.008$\\
 9  & 0.10 &  (7.07 , 0.00)    & 0.027 & $0.0554 \pm 0.0002$ & $0.0624 \pm 0.0011$ & $0.766 \pm 0.002$ & $0.716 \pm 0.007$ & $0.138 \pm 0.006$\\
10  & 0.05 & (10.95 , 0.00)    & 0.044 & $0.0278 \pm 0.0001$ & $0.0258 \pm 0.0004$ & $0.731 \pm 0.003$ & $0.644 \pm 0.005$ & $0.232 \pm 0.006$\\
11a & 0.06 &  (8.00 , 6.00)    & 0.037 & $0.0332 \pm 0.0002$ & $0.0299 \pm 0.0021$ & $0.755 \pm 0.005$ & $0.654 \pm 0.003$ & $0.163 \pm 0.005$\\
11b & 0.06 & (10.00 , 0.00)    & 0.037 & $0.0337 \pm 0.0002$ & $0.0317 \pm 0.0006$ & $0.724 \pm 0.005$ & $0.650 \pm 0.007$ & $0.194 \pm 0.005$\\
12  & 0.08 &  (9.35 , 0.00)    & 0.022 & $0.0479 \pm 0.0004$ & $0.0479 \pm 0.0014$ & $0.653 \pm 0.004$ & $0.630 \pm 0.004$ & $0.112 \pm 0.012$\\
13  & 0.09 &  (9.43 , 0.00)    & 0.017 & $0.0561 \pm 0.0001$ & $0.0553 \pm 0.0003$ & $0.589 \pm 0.002$ & $0.558 \pm 0.003$ & $0.086 \pm 0.005$\\
14  & 1.00 &  (3.00 , 0.00)    & 0.001 & $0.6597 \pm 0.0001$ & $0.6494 \pm 0.0007$ & $0.514 \pm 0.001$ & $0.501 \pm 0.001$ & $0.020 \pm 0.006$\\
\hline
S2&  -   &        -          & 0.0007& $4.6256 \pm 0.0053$ & $4.6140 \pm 0.0317$ & $0.892 \pm 0.002$ & $0.888 \pm 0.003$ & $0.002 \pm 0.005$\\
\hline
\end{tabular}
\end{table*}

\begin{figure}[htbp]
\centering
\subfigure{\includegraphics[width=0.49\textwidth]{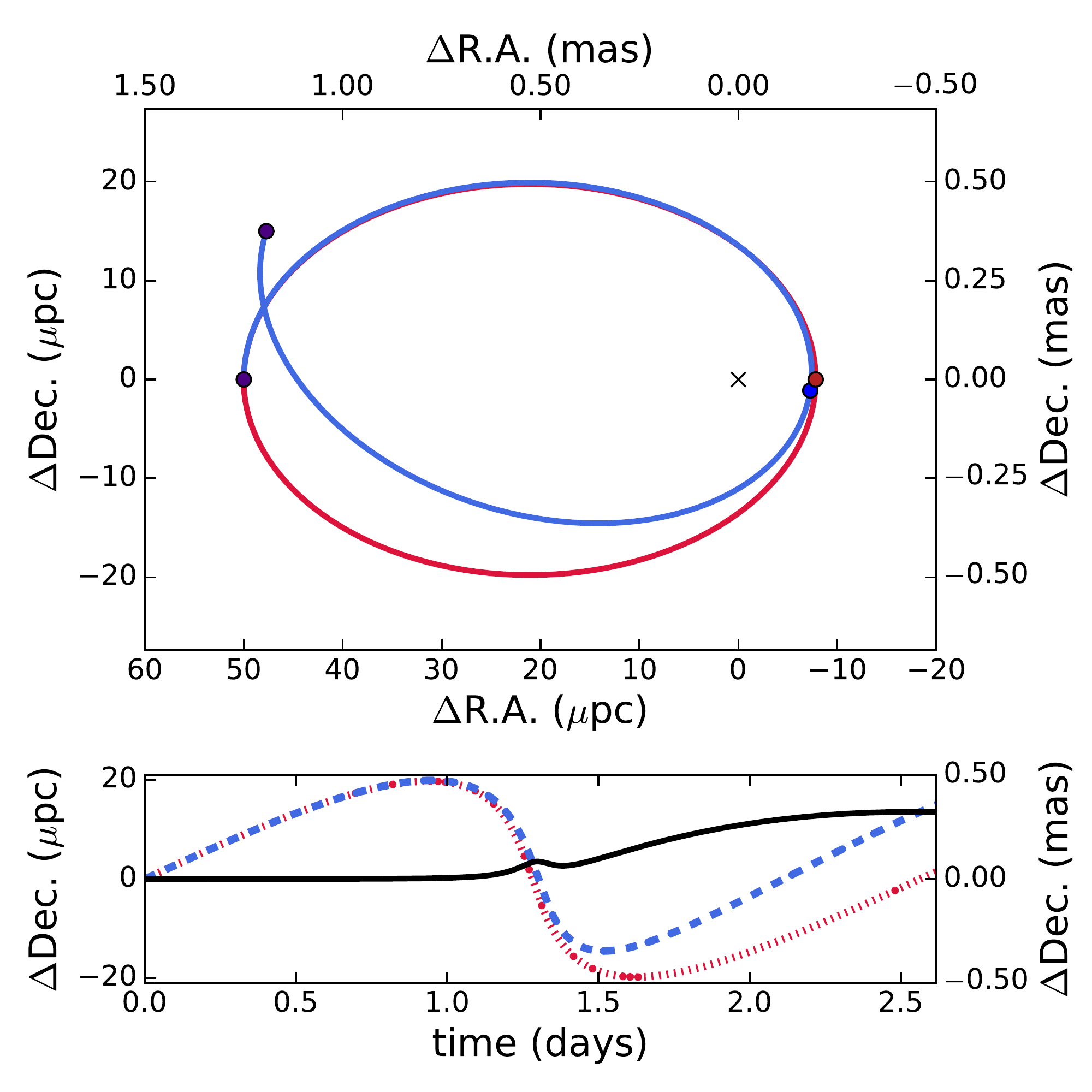}}
\caption{Comparison between the Newtonian and the relativistic orbits (case 7a of Table~\ref{table:noi}). The top panel shows the relativistic/Newtonian orbit as a blue/red line. The blue/red circle shows the periapse position of the relativistic/Newtonian orbit. The purple points mark the apoapses. The bottom panel shows the declination ($\Delta$Dec.) of the two orbits and the difference between them against time, because (as a result of the orientation of the orbit) most of the deviation from a Newtonian orbit happens in this direction. The dashed blue line is the $\Delta$Dec. of the relativistic orbit, and the dotted red line is the $\Delta$Dec. of the Newtonian  orbit and the solid black line is the difference between them. Note that the peak is at the periapse.}
\label{fig:relkeporb}
\end{figure}

\indent Figure~\ref{fig:relkeporb} demonstrates two different paths a star (case 7a of Table~\ref{table:noi}) will take with the same initial position and momentum, on the sky plane, when the orbit is purely Newtonian (red line) and when the first-order PN approximation is used. The blue/red circle shows the periapse position of the relativistic/Newtonian orbit. The purple points mark the positions of the apoapses for both orbits. For a Newtonian orbit the apoapse positions overlap, while for a relativistic orbit there is a shift in the apoapse position after one orbit. The orbit is oriented horizontally so most of the difference between the positions can be seen in the declination ($\Delta$Dec.) direction. The $\Delta$Dec. of the two orbits, the dashed blue line for the relativistic orbit and the dotted red line for the Newtonian, and the difference between them, the solid black line are plotted against time in the bottom panel. As a result of the periapse shift, most of the deviation from the Newtonian orbit happens after the periapse. There is a peak in the plot, which indicates the periapse.

\begin{figure}[htbp]
\centering
\subfigure{\includegraphics[width=0.49\textwidth]{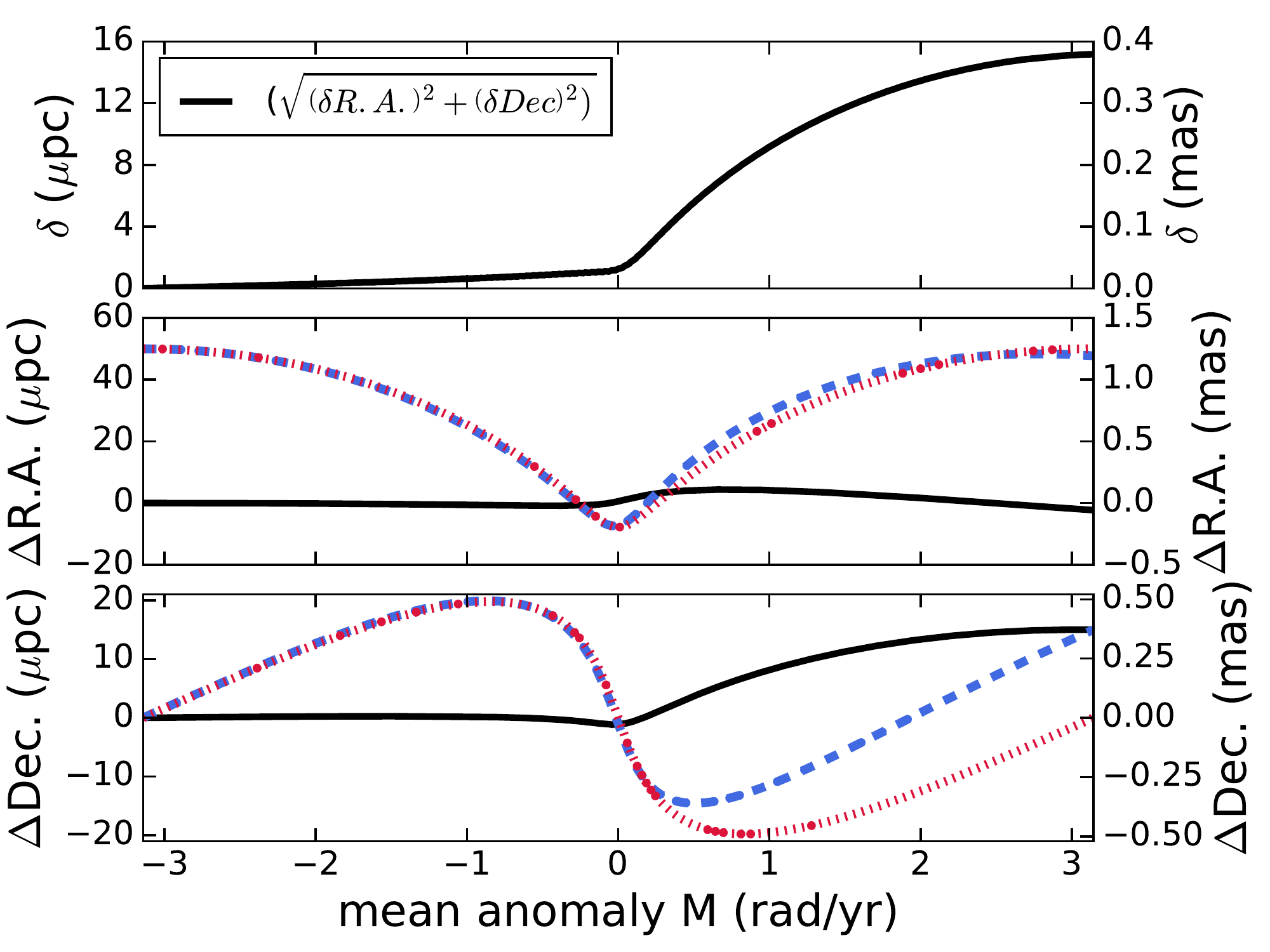}}
\caption{Comparison between the Newtonian and the relativistic orbits (case 7a of Table~\ref{table:noi}) as a function of orbital phase. The middle and bottom panels show the $\Delta$R.A. and $\Delta$Dec. of the orbits demonstrated in Fig. \ref{fig:relkeporb} with respect to their mean anomaly in units of rad yr$^{-1}$. The zero mean anomaly ($M$) is the periapse and $-\pi$ and $\pi$ are the first and second apoapses. The relativistic orbit is dashed blue, and the Newtonian one is dotted red, and the difference between the two in each panel is shown with a solid black line. The black solid line in the first panel shows $\sqrt{\left(\delta R.A.\right)^2+\left(\delta Dec\right)^2)}$, which is the distance between the positions of the two cases.}
\label{fig:delta}
\end{figure}

\indent However, if one wants to compare the orbits at the same phase one should plot the positions against the mean anomaly. The middle and bottom panels in Fig.~\ref{fig:delta} show the $\Delta$R.A. and $\Delta$Dec. of the relativistic orbit with a dashed blue line and those of the Newtonian with a dotted red line with respect to their mean anomaly in units of rad yr$^{-1}$. The difference between the two lines in each panel is shown with a solid black line. The top panel depicts the distance between the positions of the two cases, defined as $\delta = \sqrt{\left(\delta R.A.\right)^2 + \left(\delta Dec.\right)^2}$. From one apoapse to the next the mean anomaly changes in the range $-\pi~<$ M $<~\pi$. The periapse happens at $M = 0$. There is no peak in the plot showing the difference in the $\Delta$Dec., since the two periapses happen at the same mean anomaly but not at the same time.

\indent As mentioned above, since all our simulated orbits are located inside the orbit of S2, the upper limit for selecting the simulated case studies is the apoapse distance of S2 ($r_a \simeq$ 0$\arcsec$.234). We disregard the orbits with periapse distances smaller than the tidal disruption radius near Sgr~A*, since main-sequence stars such as S-stars cannot exist within this radius. The tidal radius for the stars in the GC is defined as $r_t \sim R_{\star}\left(M_{BH}/M_{\star}\right)^{(1/3)} \simeq$ 85~$\mu$as, given by \citet{alex2005} for a star of mass $M_{\star}= 1$~\msun, a radius of $R_{\star} = 1$~$R_{\odot}$, and a SMBH mass of $M_{BH}$~= 3.5~$\times$~10$^6$~\msun. The mass of the SMBH is estimated to be $\sim$4.3~$\times$~10$^6$~\msun\ in this study, thus the tidal radius is $r_t \sim$~90~$\mu$as.

\begin{figure}[htbp]
\centering
\subfigure{\includegraphics[width=0.49\textwidth]{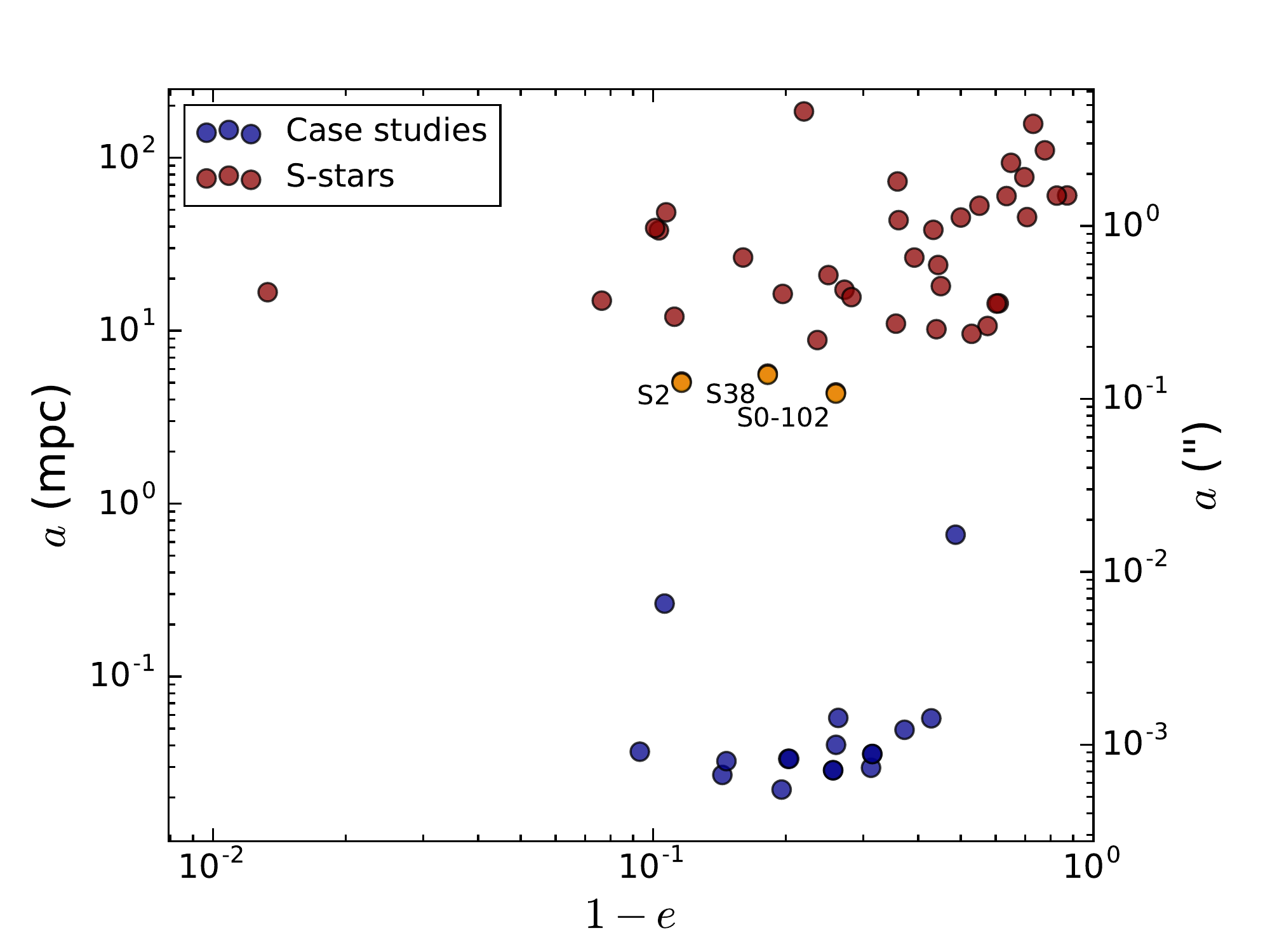}\label{fig:dis_a_e}}
\subfigure{\includegraphics[width=0.49\textwidth]{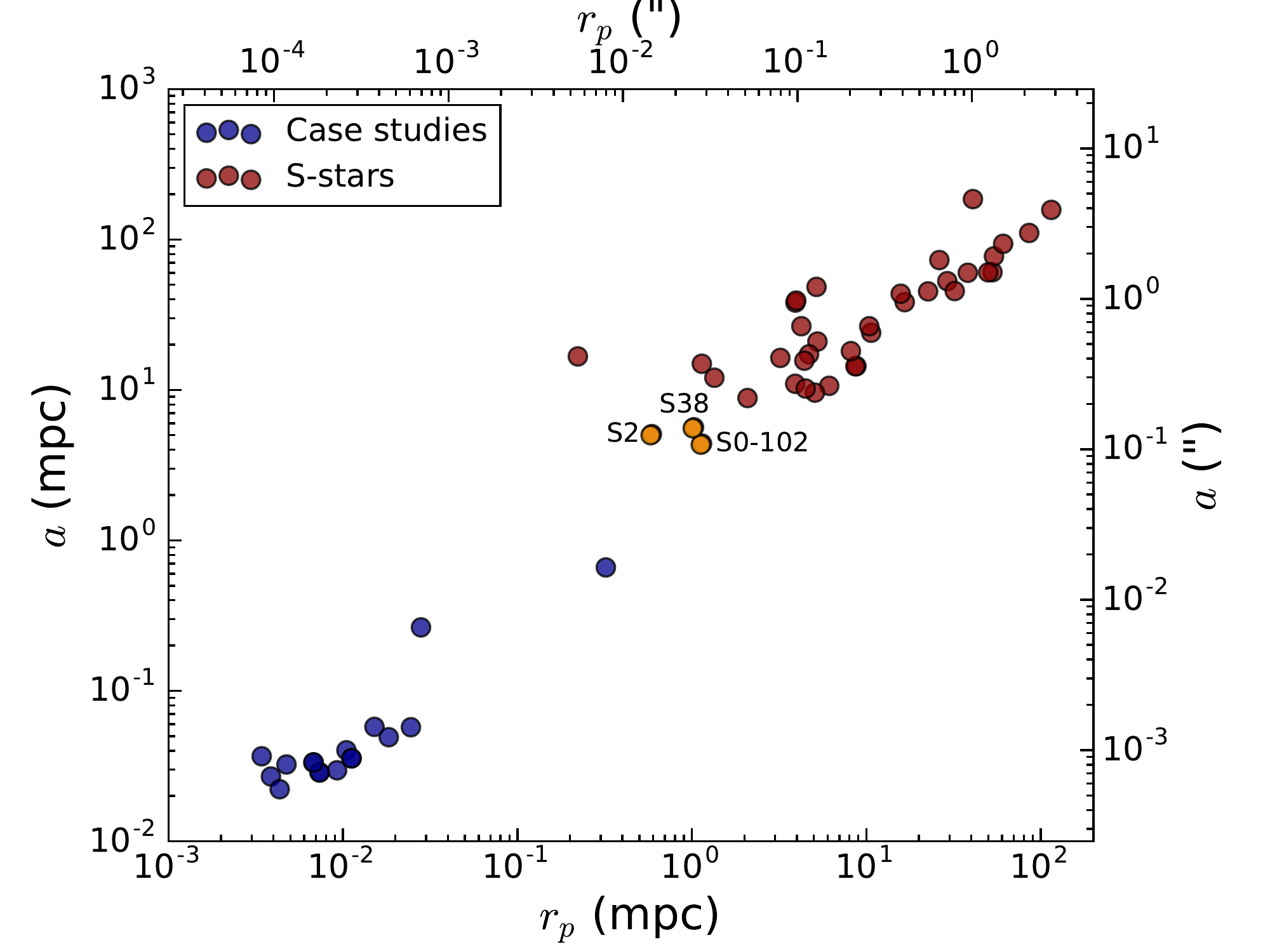}\label{fig:dis_a_rp}}
\caption{Distribution of the S-stars with determined orbits (red circles) and the possible cases that might exist closer to the SMBH that are presented here and listed in Table~\ref{table:noi} (blue circles). In the top panel, we plot semimajor axis against eccentricity and in the bottom panel, we plot semimajor axis against periapse distance. S111 is not shown here since its orbit is hyperbolic. S2, S38, and S55/S0-102 are shown with orange circles and are labeled. The semimajor axis and the eccentricity of all the S-stars except S2, S38, and S55/S0-102 have been taken from \citet[][Table 3]{gill17}.}
\label{fig:distribution}
\end{figure}

\indent The distribution of these simulated stars compared to the distribution of the detected S-stars is depicted in Fig.~\ref{fig:distribution}. The S-stars are shown with red circles while the simulated stars are shown with blue circles in the plots of semimajor axis versus eccentricity (top panel) and semimajor axis versus periapse distance (bottom panel). S2, S38, and S55/S0-102 are shown with orange circles and labelled.

\indent The stars for our highly relativistic case studies cover a similar range in eccentricity 
to the S-stars but smaller orbital axes. With this coverage we start developing a method to estimate and predict how relativistic the orbits of stars are after observing them for just one orbit using only astronomical data. We should keep in mind that our main goal is to find a correlation between the theoretical relativistic and the observational parameters.

\subsection{Developing methods to measure the PN effects}
\label{sec:method}

\indent To develope a method to measure the strength of the PN effects, first we have to find an observable that is relatively easy to measure and that changes noticeably as a result of these effects. In the search for such an observable, which also enables us to exploit the uncertainties in the orbital fitting, we employ a concept similar 
to that of the squeezed states.

\subsubsection{Squeezed States}
\label{sec:sstate}

\indent In our framework of orbital fitting we consider a squeezed state to be any state in which an uncertainty principle is fulfilled and saturated. As a precondition for this, we need the product of two quantities $\alpha$ and $\beta$ to be larger than but very close to a minimum limit 

\begin{equation}
\label{eqn:sqs1}
\alpha \times \beta \geq \epsilon~~. 
\end{equation}

In this context the term uncertainty implies that the fitting procedure allows a considerable portion of the measurement uncertainties to be shifted back and forth between the two quantities. Then we consider the result of the fitting procedure to be a state that allows us to squeeze the uncertainties into one of the variables.
   
\indent If we set $\alpha$ = $e^{-\chi^{2}_{l}}$ and $\beta$ = $e^{-\chi^{2}_{u}}$ with $\chi^{2}$ being the weighted sum of the squared errors as a result of fitting an elliptical orbit to the observational data, then e$^{-\chi^{2}}$ describes the goodness of the fit or its likelihood of representing the orbit well. The subscripts \textit{u} and \textit{l} in this paper denote the upper and lower parts of the orbit respectively. Hence, we write

\begin{equation}
\label{eqn:sqs2}
{e}^{-\chi^{2}_{l}} \times {e}^{-\chi^{2}_{u}} \geq {e}^{-\chi^{2}}
\end{equation}

\noindent or

\begin{equation}
\label{eqn:sqs3}
{\chi}^{2}_{l}+{\chi}^{2}_{u} \geq \chi^{2}.
\end{equation}

\indent We can now separate the fitting errors contributed by the random uncertainties of the measurement (subscript \textit{r}) and those contributed by the misfit of the orbital shape if it is not a perfect ellipse (subscript \textit{s}). The inequality \eqref{eqn:sqs3} is true (i.e. the inequality sign is justified) if we set $\chi = {\chi}_{lu,r}$, which describes the overall goodness of the combined upper and lower orbital fit just based on the random measurement uncertainties, so

\begin{equation}
\label{eqn:sqs4}
\chi^{2}_{l,s} + \chi^{2}_{u,s} + \chi^{2}_{l,r} + \chi^{2}_{u,r} \geq \chi^{2}_{ul,r}.
\end{equation}

\indent The inequality is largely compensated for if we include on the right side of the equation the goodness of the combined upper and lower orbital fit just based on the mismatch of the overall orbital shape:

\begin{equation}
\label{eqn:sqs5}
\chi^{2}_{l,s} + \chi^{2}_{u,s} + \chi^{2}_{l,r} + \chi^{2}_{u,r} \sim \chi^{2}_{ul,s} + \chi^{2}_{ul,r}.
\end{equation}

\indent If we assume that the measurement uncertainties for the upper and lower parts of the orbit are similar, $\chi^{2}_{u,r} \sim \chi^{2}_{l,r} $, then we can write

\begin{equation}
\label{eqn_sqs6}
\chi^{2}_{lu,r} = \chi^{2}_{l,r} + \chi^{2}_{u,r} \sim 2 \times \chi^{2}_{l,r}.
\end{equation}

For a global fit to the entire orbit the uncertainties are distributed 
well between the upper and lower parts of the orbit, and we 
assume that the measurement uncertainties are similar, 
$\chi^{2}_{u,s} \sim \chi^{2}_{l,s} $:

\begin{equation}	
\label{eqn:sqs7}
\chi^{2}_{lu,s} = \chi^{2}_{l,s} + \chi^{2}_{u,s} \sim 2 \times \chi^{2}_{l,s} \sim 2 \times \chi^{2}_{u,s}.
\end{equation}

\begin{figure}[htbp]
\centering
\includegraphics[width=0.5\textwidth]{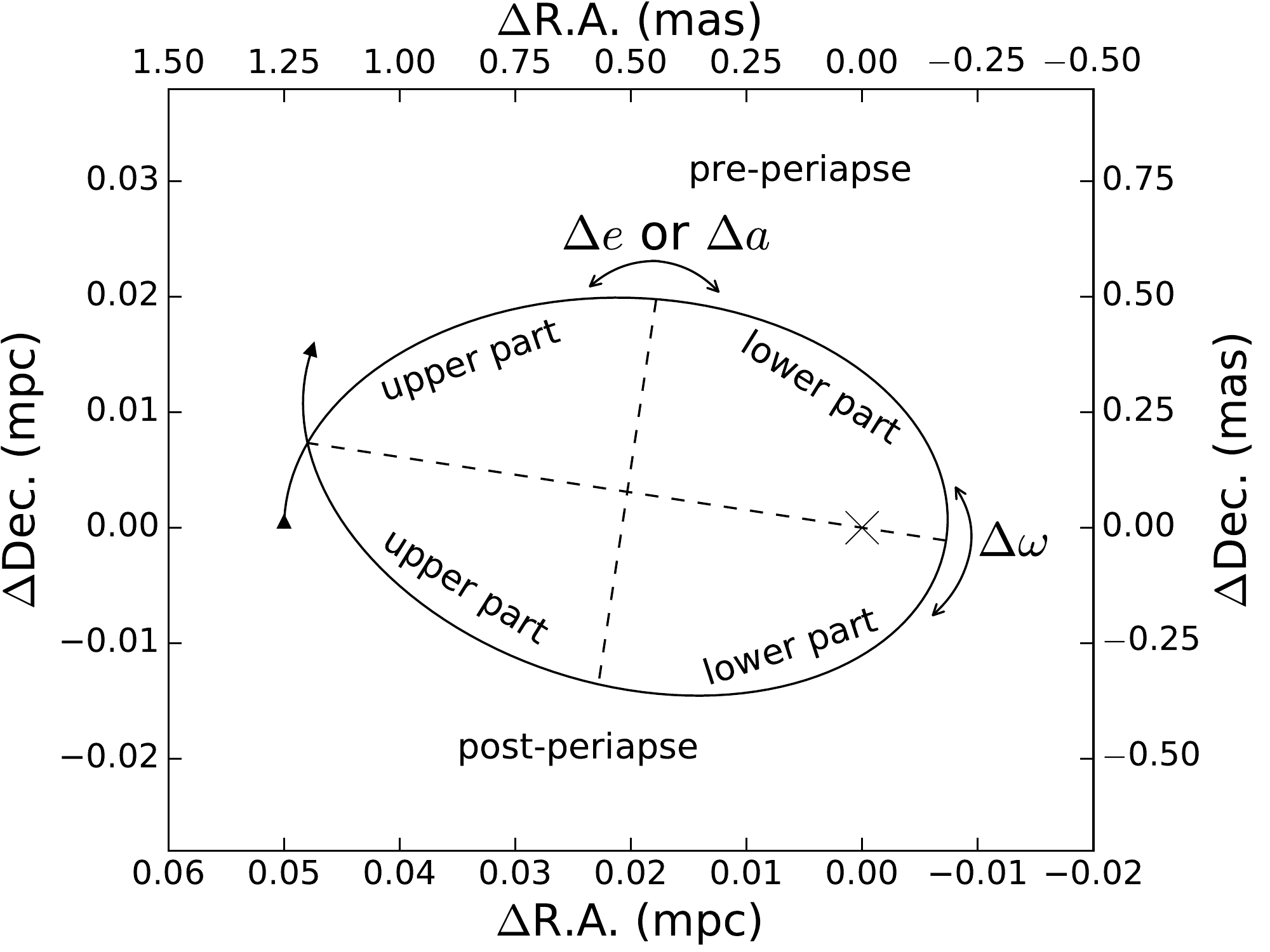}
\caption{Comparison between the methods described in Sect. \ref{sec:sstate} for observation of the relativistic effects by means of measuring the changes in the orbital parameters in two halves of the orbit.}
\label{fig:compare}
\end{figure}

\indent 
However, if we exclusively fit the lower or upper part of the orbit well, 
the situation changes.
If the measurements are very precise then on one side the uncertainties
due to the mismatch of the shape may become dominant  
and much larger than the random uncertainties.
As can be seen in Figs.~\ref{fig:compare} and \ref{fig:method} this 
may be relevant the for relativistic orbits. In this case, 
the ellipses fitted to the lower part of the orbit have 
systematically lower ellipticities and semimajor axes.
If fitting one side only, the uncertainties due to a mismatch of 
the orbital shape will be concentrated (squeezed) to the orbital 
section on the opposite side and we find

\begin{equation}
\label{eqn:sqs8}
if \hspace{4mm} \widetilde{\chi^{2}_{l,s}} \longrightarrow 1 \hspace{4mm} then \hspace{4mm} \widetilde{\chi^{2}_{u,s}} \longrightarrow \hspace{1mm} \sim 2 \times \chi^{2}_{u,s} > 1.
\end{equation}

\begin{equation}
\label{eqn:sqs9}
if \hspace{4mm} \widetilde{\chi^{2}_{u,s}} \longrightarrow 1 \hspace{4mm} then \hspace{4mm} \widetilde{\chi^{2}_{l,s}} \longrightarrow \hspace{1mm} \sim 2 \times \chi^{2}_{l,s} > 1.
\end{equation}

\indent Here, $\widetilde{\chi^{2}_{l,s}}$ and $\widetilde{\chi^{2}_{u,s}}$ are the squeezed 
$\chi^{2}$-values one obtains after the fit to only one side of the orbit.
Therefore, the ratio between $\chi^{2}_{u,s}$ or $\chi^{2}_{l,s}$ and these two quantities can be used to decide on
the degree to which the orbit is dominated by relativistic effects. Figure \ref{fig:chisqr} demonstrates the correlation between \ups~and one of these ratios ($\chi^{2}_{u,s}/\widetilde{\chi^{2}_{u,s}}$) for the cases in Table \ref{table:noi}. If $\chi^{2}_{u,s} \sim \chi^{2}_{l,s} $ then we expect 1/2 for the ratio (equation \eqref{eqn:sqs8}), but for more relativistic cases $\widetilde{\chi^{2}_{u,s}}$ gets even larger than $2 \times \chi^{2}_{u,s}$. As we go to more Newtonian cases, the ratio approaches unity since the systematic differences between the upper and the lower parts of the orbit disappear. The best fit describing the correlation is

\begin{equation}
\label{eqn:upschi}
\chi^{2}_{u,s}/\widetilde{\chi^{2}_{u,s}} = e^{(-16.23 \pm 0.13) \Upsilon}.
\end{equation}

\noindent However, squeezing the goodness of fit to only one side of the orbit
has consequences for the orbital elements. 
For an ellipse we know

\begin{equation}
\label{eqn:sqs10}
e = \sqrt{1-\left(\frac{b}{a}\right)^{2}},
\end{equation}

\noindent where $\mathnormal{e}$ is the eccentricity, $\mathnormal{a}$ is the semimajor axis,and $\mathnormal{b}$ is the semiminor axis of the ellipse. Equation \eqref{eqn:sqs10} shows that deviations of the overall orbital 
shape from an ellipse will become apparent in the misfit of the 
semimajor axis and eccentricity if we fit only to the lower or upper 
half of the orbit, respectively.
Such deviations from Newtonian ellipses are expected if the orbits 
are influenced by relativistic effects or by an additional 
smooth or granular extended mass distribution.
The differences can be expressed as ratios of 
parameters $a_l/a_u$ and $e_l/e_u$.
Here, the subscripts again indicate for which orbital section the fit
was optimized.

\indent The method we present can be compared to a method by \cite{AngelilSaha2014} to show the relativistic effects on the argument of the periapse $\omega$. The predicted pericenter shift for S2 during its Sgr~A* flyby in 2018
will amount to about 11$'$ (for a semimajor axis of 0\arcsec.126, an eccentricity of 0.88, and a BH mass of 4.15~$\times$~10$^6$~\msun), but \cite{AngelilSaha2014} point out that it does not occur smoothly. They find that the difference in $\omega$ from the 
pre- to post-periapse part of the orbit occurs almost in a step at each pericenter passage. This method is basically equivalent to a measurement of the periapse shift using the two halves of the orbit before and after the periapse.

\indent With the ratios $a_l/a_u$ and $e_l/e_u$ our method uses the orbital differences in the radial direction (see Fig. \ref{fig:compare})
and we exploit the folding symmetry along the semiminor axis (as expected from a Newtonian motion) while \cite{AngelilSaha2014} make 
use of the folding symmetry of the orbit along the semimajor axis. 
Hence, their method corresponds to squeezing the goodness of the orbital fit to the pre-/post-periapse part of the orbit (Fig.~\ref{fig:compare}) and therefore can also be described using the formalism we present here.

\indent The rapid change of $\omega$ can in principle be derived using only small sections of the pre-/post-periapse part of the orbit.
However, one still needs the information on the full orbit, because the orbital solutions need to be derived in the same orbital plane 
in order to represent the progressing periapse shift precisely. 
Using elliptical fits to all or most of the upper/lower or pre-/post-periapse part of the orbit, we ensure that the entire orbit 
within its plane is represented in as complete a way as possible.

\subsubsection{Comparison of Methods}
\label{sec:method1&2}

\begin{figure*}[htbp]
\centering
\subfigure{\includegraphics[width=0.31\textwidth]{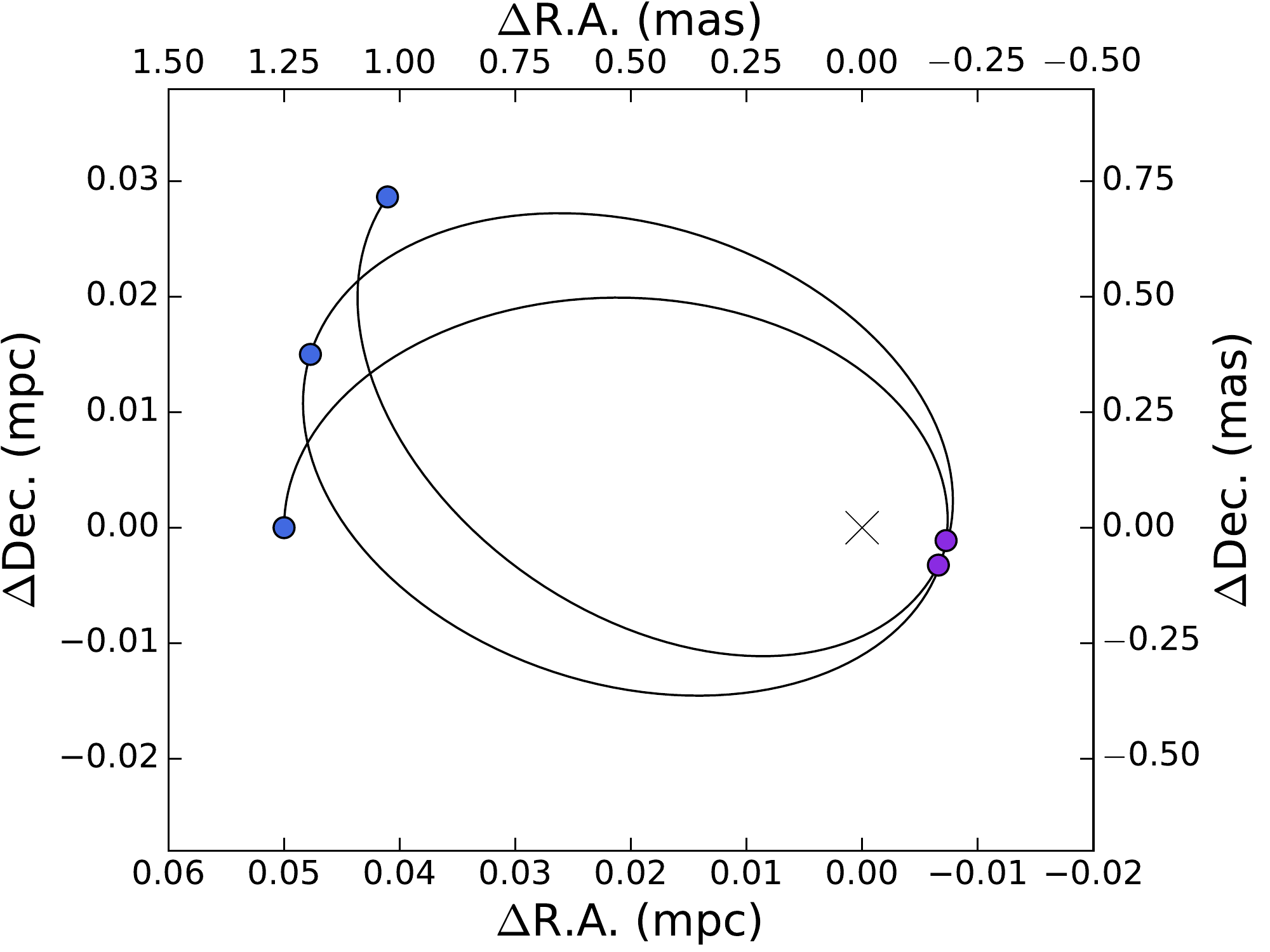}\label{fig:relorb}}
\subfigure{\includegraphics[width=0.31\textwidth]{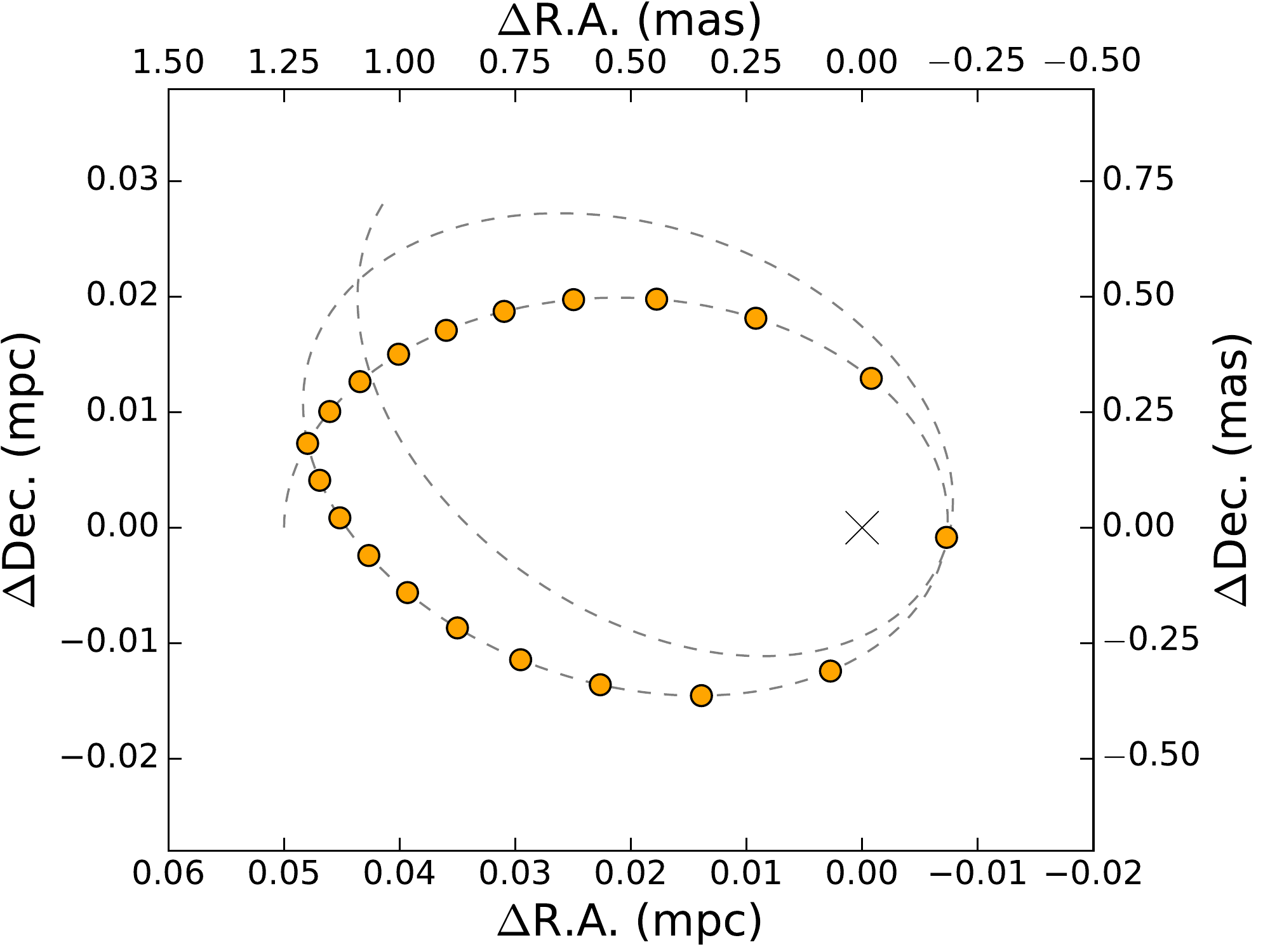}\label{fig:narrowed}}
\subfigure{\includegraphics[width=0.31\textwidth]{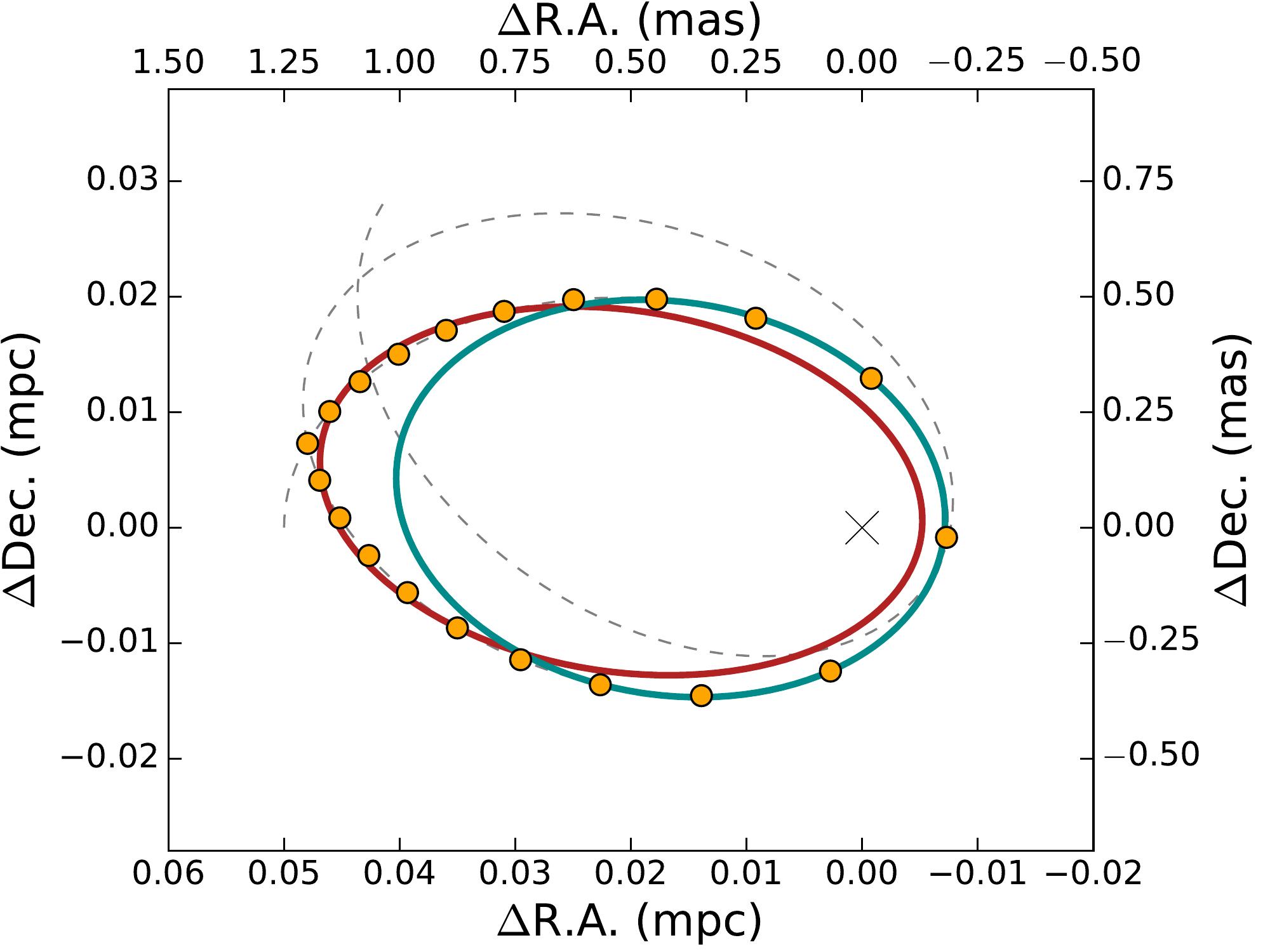}\label{fig:fits}}
\caption{Example of the method described in section~\ref{sec:sim}. Left panel: two full orbits of the example star with the blue circles marking the three apoapses and the violet circles marking the two periapses. These orbits are also shown with dashed lines in the middle and right hand panels.
Middle panel: The observed data points on the sky plane shown as orange circles are considered to be part of a closed orbit. The observations are assumed to have taken place after equal time intervals. The data points have larger separations around the periapse due to the higher velocity than in the rest of the orbit. 
Right panel: the red curve shows the elliptical fit to the upper half of the observed data points and the cyan curve shows the elliptical fit to the lower half. The location of the SMBH at the center is marked with a black cross in all panels.}
\label{fig:method}
\end{figure*}

\begin{figure}[htbp]
\centering
\includegraphics[width=0.5\textwidth]{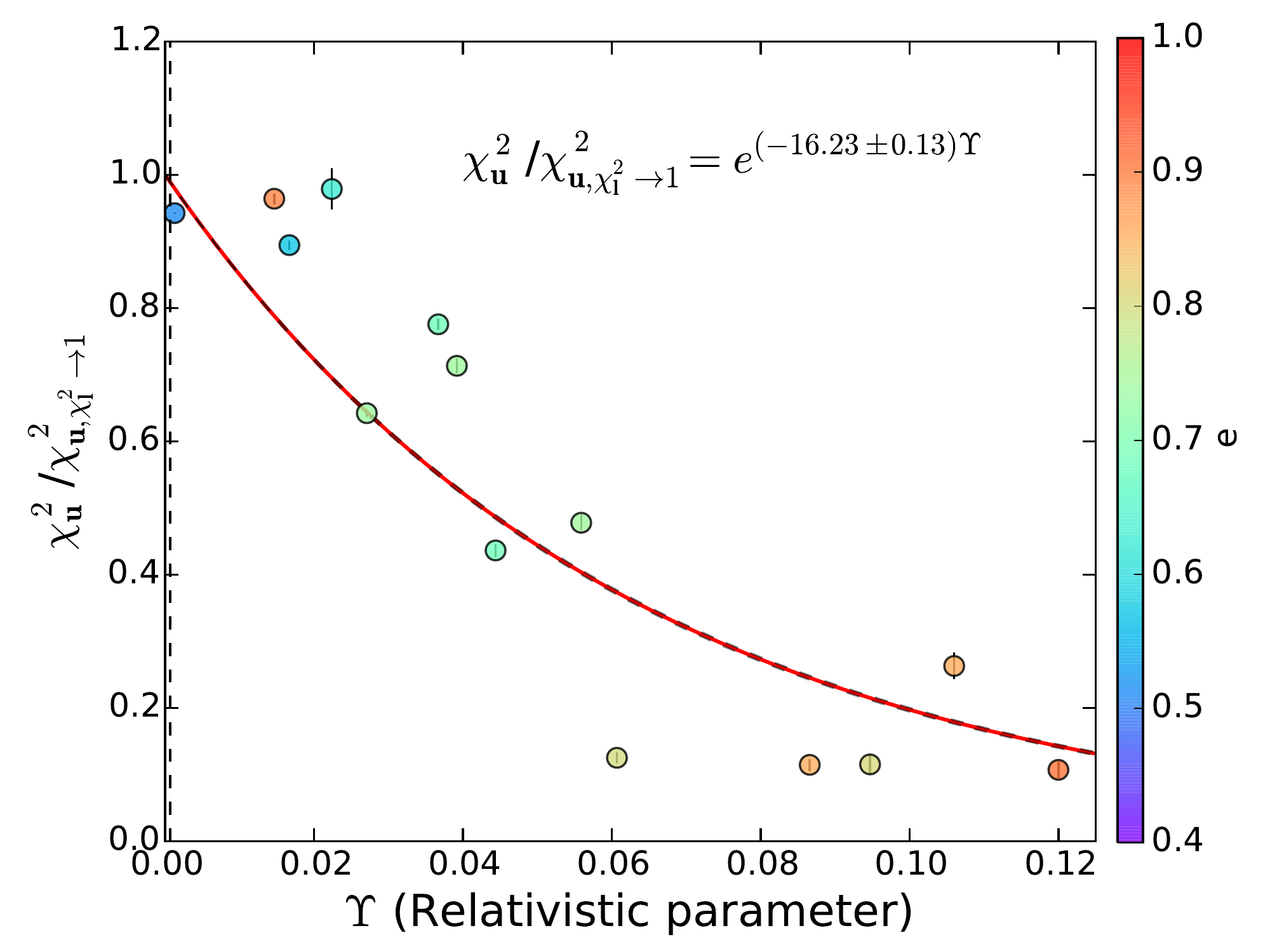}
\caption{Correlation between the relativistic parameter \ups\ and the ratio of the goodness of fit to the upper part of the orbit to the squeezed one after fitting only to the lower part $\chi^{2}_{u}/\chi^{2}_{u,\chi^{2}_{l} \to 1}$ for the case studies in Table \ref{table:noi}.}
\label{fig:chisqr}
\end{figure}

\indent In order to illustrate the method of $a_l/a_u$ and $e_l/e_u$ ratios and the $\Delta \omega$ method,
we consider two full orbits of the case 7b from Table~\ref{table:noi} (see Fig.~\ref{fig:method}).
The apoapses and periapses are marked by blue and violet circles respectively in the left panel. 
The simulated data points shown with yellow circles make a full closed orbit on the sky in the middle panel.
These data points are generated after equal time intervals.
We fit an ellipse to the upper half of the data, shown as a red solid ellipse in the right panel. 
Subsequently, we fit another ellipse to the lower half of the data, shown as a cyan solid ellipse in the same panel.

\indent We repeat these steps for all our case studies and report the different values for the semimajor axis and eccentricity of the two halves in columns 5--8 of Table~\ref{table:noi}. The elliptical orbits and the orbital elements are obtained by a $\chi^{2}$ fit to the data points using the L-BFGS-B (limited-memory Broyden--Fletcher--Goldfarb--Shanno) minimization method. In order to estimate a value for the uncertainty of the measurement of each data point in the fits, we use the standard deviation of all the upper or lower half of the data points considering no mean displacement from a Newtonian orbit for half of the orbit, and an approximate mean displacement of $\mathnormal{a}$$\Delta\omega$/4 for the other half. Here $\Delta\omega$ is the periapse shift. 
The uncertainties include misfitting ellipses to the upper and lower parts of the orbit. The time intervals between the data points are in accordance with the scale of the orbit and consequently its orbital period, which can be between an hour and a month.

\begin{figure}[htbp]
\centering
\subfigure{\includegraphics[width=0.49\textwidth]{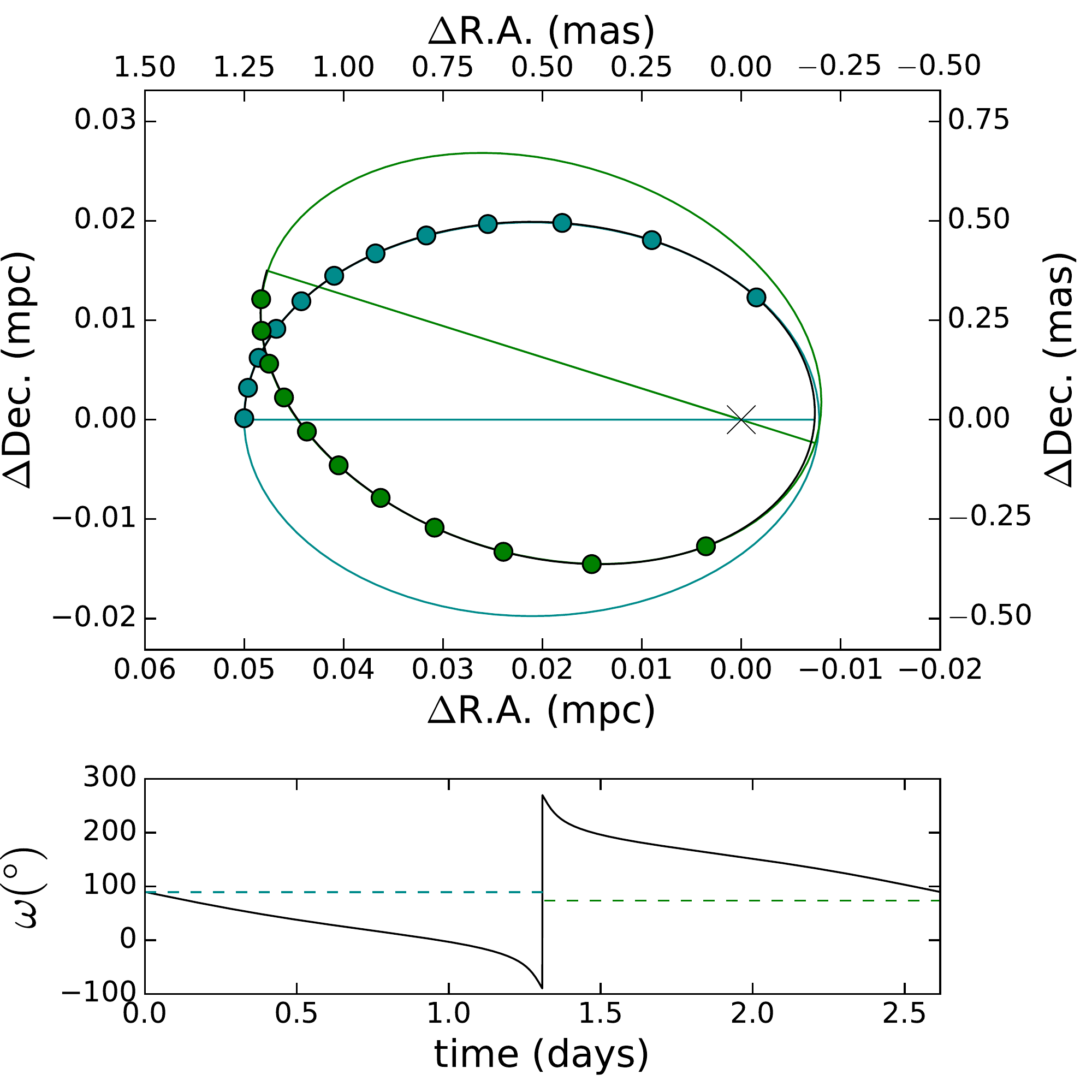}\label{fig:instomega}}
\caption{Method by \cite{AngelilSaha2014} for the observation of the relativistic effects by means of measuring the changes in the argument of the periapse $\omega$ in two halves of the orbit, applied here to the case 7b from Table~\ref{table:noi}. Upper panel: the data points from two halves of the orbit, before and after the periapse, and their fits are shown by cyan and green lines, respectively. The periapse shift is the angle between the two major axes. Bottom panel: the instantaneous argument $\omega$ of periapse for one period. The cyan and green dashed lines are the argument of the periapse of the fit to the respective half of the orbit. The instantaneous argument of the periapse $\omega_{inst}$ at each point of an orbit is the argument of periapse of a Newtonian orbit with the same position and momentum at that specific point. The quantity $\omega_{inst}$ is not an observable.}
\label{fig:method2}
\end{figure}

\indent Next, we apply the $\Delta\omega$ method discussed in Sect.~\ref{sec:sstate} by fitting ellipses to the pre- and post-periapse halves of the orbits of the case studies. This method is illustrated for the case 7b from Table \ref{table:noi} in Fig \ref{fig:method2}. We get the same values for all the orbital elements except the argument of the periapse for the two fits, as expected. The difference in $\omega$ of the two halves for each case study is reported in the last column of Table \ref{table:noi}.

\indent Without loss of generality all our case studies are located on the sky plane, i.e. they have no velocity component in the $z$ direction. For the sake of completeness, we add a $z$ velocity component to three of the case studies and consequently generate three cases with inclination (6a, 7a, and 11a in Table~\ref{table:noi}). Applying the two methods to these cases, assuming a real situation in which we do not have any information about the inclination but the radial velocity, includes a few more steps. Before fitting the ellipses for each of them we first have to fit a simple Newtonian orbit to the astrometric and radial velocity data to find the inclination and the line of nodes of the orbit. Then using the inclination and the argument of the periapse we have to 
correct the astrometric data for inclination. Afterwards we can continue with the rest of the steps. However, as a consequence of these additional steps, we have additional sources of uncertainty. Therefore, we choose not to use these cases for our analysis.  

\subsection{The case of S2}
\label{sec:rels2}

\indent To see how well the methods work, we apply them to the 
data of one full S2 orbit. Since our VLT data from 2002 to 2015
do not allow us to cover one full orbit and since we lack 
sufficient data on the lower half of the orbit, we add the data provided 
by \citet{boehle16} from 1995 to 2010 to the data presented in this  work. 
However, the two data sets do not share the same reference coordinate system. We use the approach discussed in \citet{gill92} to bring the two data sets into a single coordinate system by assuming that they differ only in the position of the origin and the zero velocity. Therefore, we add four additional parameters---$\Delta x$, $\Delta y$, $\Delta v_x$, and $\Delta v_y$---to our parameters. Also, we make use of the PN approximation and the astrometric data presented here and in \citet{boehle16}, and the radial velocity data in the latter and in \citet{gill92}. Finally we fit simultaneously for the S2 orbital parameters, the gravitational potential parameters, and the new parameters to bring the two data sets into a single coordinate system. The fit results for these four new parameters are

\begin{align}
\label{eqn:reference}
\Delta x~ &= +2.95 \pm 0.25~\text{(mas)}\nonumber\\
\Delta y~ &= -1.08 \pm 0.48~\text{(mas)}\nonumber\\
\Delta v_x &= -0.21 \pm 0.04~\text{(mas/yr)}\nonumber\\
\Delta v_y &= -0.44 \pm 0.09~\text{(mas/yr)}.
\end{align}

These values refer to the epoch of 1995 May, which is chosen as the zero time of the orbital fit. The uncertainties are the results of MCMC simulations. Using these parameters and applying them on the data from \citet{boehle16}, we have sufficient data covering all quarters of the orbit to apply the two methods to S2. The resulting semimajor axis, eccentricities, and $\Delta \omega$ for S2 using the two methods are in the last row of Table~\ref{table:noi}.

\begin{figure}[htbp]
\centering
\subfigure{\includegraphics[width=0.49\textwidth]{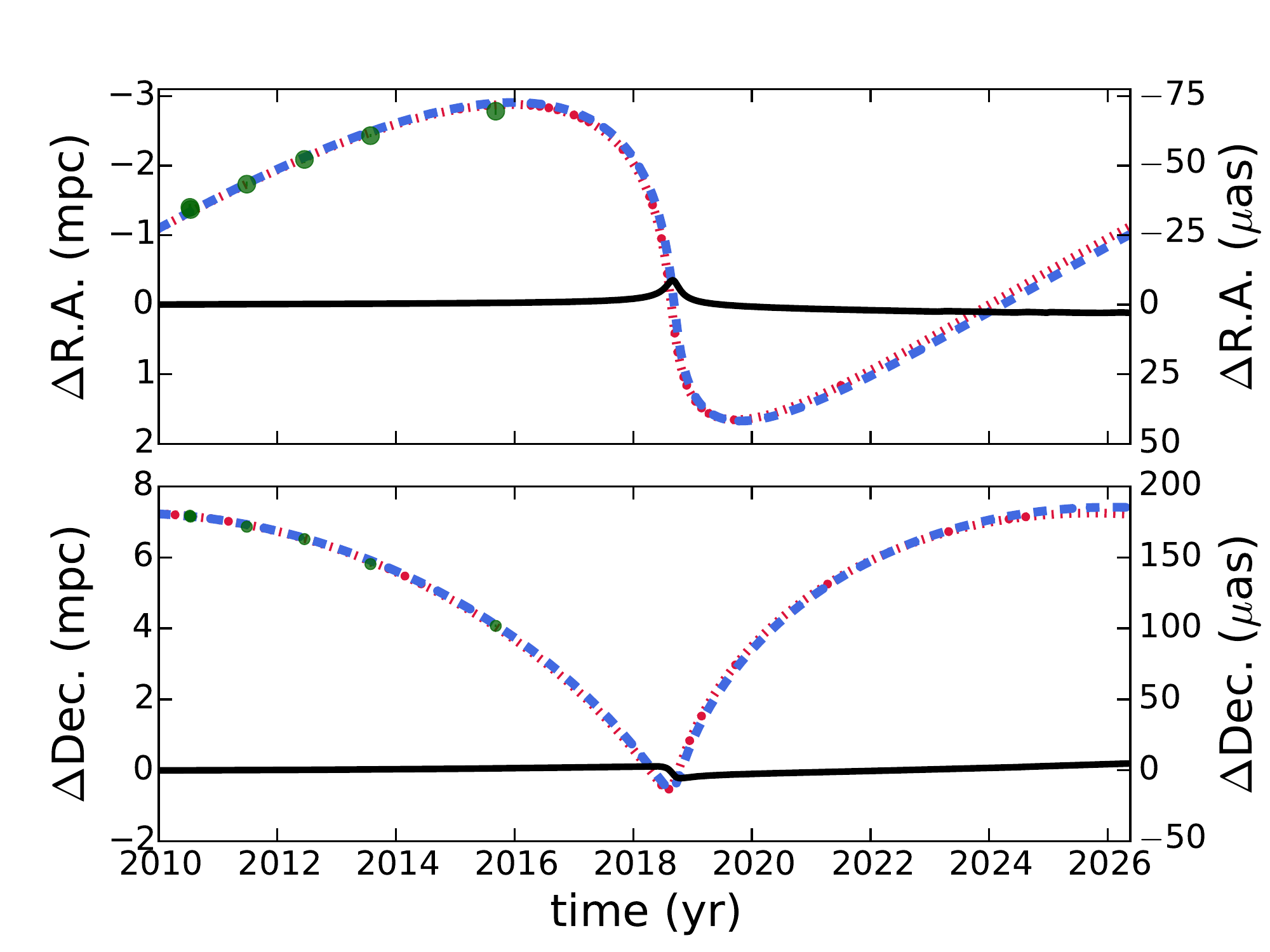}}
\caption{Comparison between the relativistic and the Newtonian fits to the astrometric and the radial velocity data of S2 from 1995 to 2015. 
The top panel shows the right ascension ($\Delta$R.A.) and the bottom panel shows the declination ($\Delta$Dec.) vs. time. The time covers one period of the orbit. 
The green circles are the data points within this period. 
The blue dashed lines are the relativistic fit. The dotted red lines represents the Newtonian fit. Both fits include the models predictions until the next periapse and beyond. 
As expected, most of the differences (shown with black solid lines) between the two orbits occur after the periapse. The peaks in the solid black lines indicate the periapse.}
\label{fig:deltas2}
\end{figure}

\indent Figure \ref{fig:deltas2} compares the Newtonian (dotted red) and relativistic (dashed blue) fits to the combined data of S2 from 1995 to 2015. It shows the difference between the two models (black solid line) in the current period and indicates that the difference will manifest itself especially during and after the periapse in 2018.

\indent Although \citet{gill92} reject the probability of a rotation between 
the VLT and Keck data sets and we follow their suggested approach in 
Sect. \ref{sec:rels2} for bringing the two data sets into one coordinate 
system, adding a parameter for rotation in our calculation in order to put 
a limit on a possible rotation is not entirely unjustifiable. 
Such a rotation (if it existed) can have an undesirable effect particularly 
on the derivation of $\Delta \omega$ of S2 in Sect. \ref{sec:rels2} for the 
reason that we used only the data provided by the Keck data set for 
the pre-periapse fit. As a result we performed a separate MCMC simulation
in which we introduced a parameter $\theta$ in addition to our four 
initial parameters 
($\Delta x$, $\Delta y$, $\Delta v_x$, $\Delta v_y$)
for describing the difference in the coordinate systems.
Without loss of generality we implemented the center of rotation 
at the location of Sgr A* and looked for the possible value 
of $\theta$ for a Newtonian orbital fit to the combined S38 
data using the data provided in \citet{boehle16} from 2004 to 2013.
The perpendicular orientation with respect to S2 on the sky 
makes S38 ideal for testing for a rotation between the two data sets. 
Irrespective of the expected small periapse shift of the orbit of S38 of
only about 6$'$, this source is ideally suited to probe relative rotations 
of the data set with respect to each other because the highest quality comparison
data are all on the northern side, i.e., on a single side, of the orbit.
We find an upper limit for the 
rotation of 0$'$.002 for the S38 orbital fit. If we repeat the procedure 
for the combination of the S2 and S38 data 
(again only on the sections of the orbits covered by both 
data sets) we find an upper limit on the rotation value of 0$'$.1. 
The expected periapse shift of S2 is about 11$'$ (for a semimajor axis of 0\arcsec.126, an eccentricity of 0.88, and a BH mass of 4.15~$\times$~10$^6$~\msun).
Since we regard the very small rotational values as upper limits we
did not apply them and we continue the analyses with the results from the MCMC 
simulations described before.
Thus we conclude that our combination of the two data sets is ideally 
suited to probing for relativistic effects on the orbital elements.

\section{Results}
\label{sec:results}

\subsection{The simulated case studies}
\label{subsec:casestudies}

\begin{figure*}[htbp]
\centering
\subfigure{\includegraphics[width=0.49\textwidth]{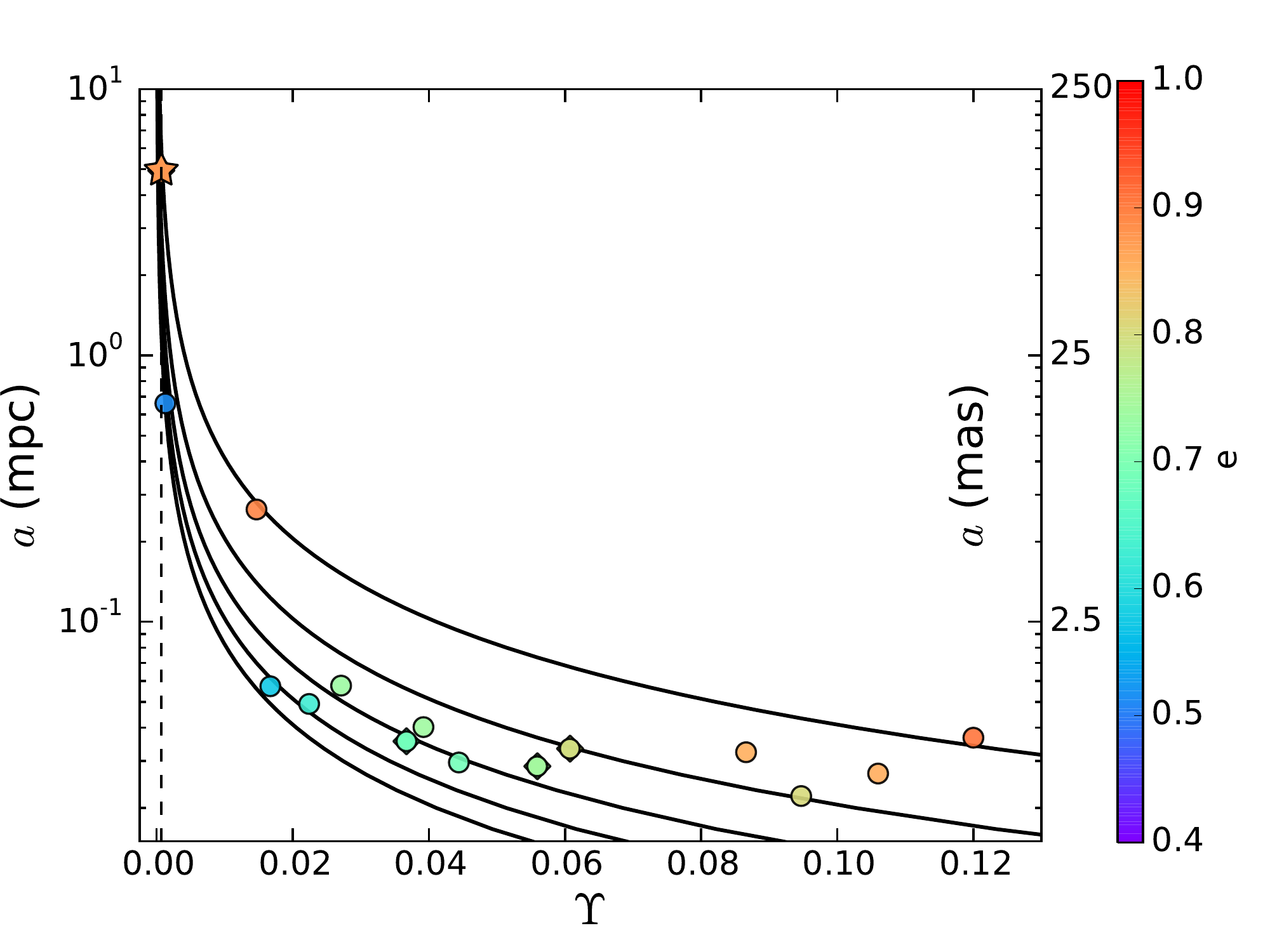}\label{fig:upsilona}}
\subfigure{\includegraphics[width=0.49\textwidth]{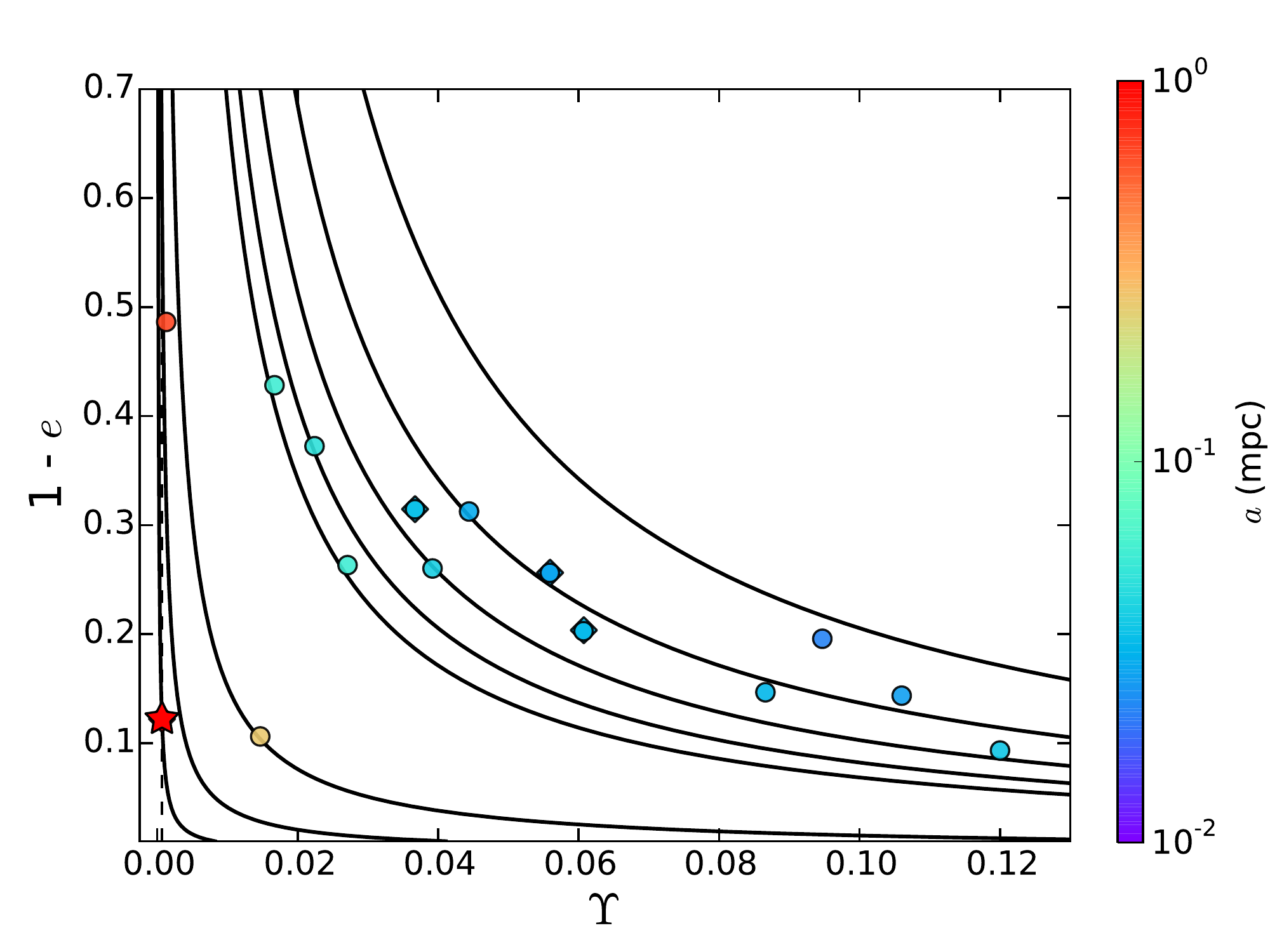}\label{fig:upsilone}}
\caption{Correlation between the relativistic parameter \ups\ and the orbital parameters, the semimajor axis in the left panel and the eccentricity via $(1 - e)$ in the right panel. The solid lines are $e =$~0.9--0.5 in steps of 0.1 from top to bottom in the left panel.
In the right panel from top to bottom the solid lines stand for $a$~=~0.02--0.06~mpc ($\sim$0.5--1.5~mas), $a =$~0.27~mpc ($\sim$6.7~ mas), $a$~=~1~mpc ($\sim$25~mas), and $a$~=~5~mpc ($\sim$125~mas).
As listed in Table \ref{table:noi} the circles represent the simulated stars on the sky plane and the diamonds show the simulated stars with (corrected for) inclinations 
with respect to the sky plane. S2 is shown with a star. The dashed line shows the lower limit for choosing the case studies, which is the expected theoretical value of \ups$_{S2}$~=~0.00065 in both panels. The color bars show the eccentricity of the orbits in the left panel and the semimajor axis in the right panel.}
\label{fig:upsilonae}
\end{figure*}

\indent A measure of the strength of the PN effects is the relativistic parameter at the periapse, $\Upsilon\equiv r_{s}/r_{p}$ \cite[see also][]{alex2005, zuc2006, ghez8}. Other suggested parameters are the dimensionless periapse distance, the periapse shift, and the speed at the periapse in units of the speed of light $\beta = v_{p}/c$, with $v_p$ being the velocity during the closest approach to the BH \citep{zuc2006}. It is a justifiable parameter for determining the approximate magnitude of the components of the Schwarzschild metric outside a single object in vacuum \citep{baker15}. $\Upsilon$ is by definition dependent on the orbital shape, i.e. the semimajor axis and eccentricity. The inverse correlations between \ups\ and the semimajor axis (and consequently the orbital period) and the eccentricity are demonstrated in Fig.~\ref{fig:upsilonae}. The solid lines are $e$~=~0.9--0.5 from top to bottom in the left panel, and $a$~=~0.02--0.06~mpc ($\sim$0.5--1.5~mas), $a$~=~0.27~mpc ($\sim$6.7~mas), $a$~= 1~mpc ($\sim$25~mas), and $a$~=~5~mpc ($\sim$~125~mas) from top to bottom in the right panel. The circles represent the simulated stars in the plane of the sky and the diamonds show the simulated stars with inclinations (corrected for) with respect to the sky plane, listed in Table~\ref{table:noi}. The dashed line shows the \ups\ of S2 for both panels, which is the minimum \ups\ in this study.

\indent The PN effects should depend only on \ups, according to the strong equivalence principle. However, this is not the case for other theories of gravitation so one should investigate the dependence of these effects on multiple parameters \citep{zuc2006}. In this study our main candidate for a parameter that can describe the strength of the PN effects in an orbit is \ups. Figure~\ref{fig:upsratiodomega} shows the correlation between \ups~and ratios of the orbital parameters $a_l/a_u$ and $e_l/e_u$, as described in Sect.~\ref{sec:cases} for the stars in Table \ref{table:noi}. 

\indent As a result of the dependence of \ups$=r_s/(a(1-e))$ on the eccentricity $e$ and the semimajor axis $a$, one might expect that in order to find a clear correlation 
between \ups\ and the ratios $a_l/a_u$ and $e_l/e_u$ (see Fig.~\ref{fig:upsilonae}), a parameterization is necessary. 
It can be understood from Fig. \ref{fig:upsilonae} and from the definition of \ups\ that the effects of $a$ and $e$ on \ups\ are only dominant on small scales and for large eccentricities.  
These effects are negligible for sufficiently large orbits with low eccentricities.
If we consider that the eccentricity of the stellar orbits typically ranges between 0.4 and 0.9 then this will result in a variation of \ups\ by a factor of 6. 
The possible semimajor axis for orbits with $r_p$ between the tidal disruption radius and $r_p$ of S2 ranges between $\sim$0.01~mpc ($\sim$0.025~mas) 
and $\sim$5~mpc ($\sim$125~mas). For changes in semimajor axis between these values, \ups\ varies by a factor of 500. 
However, in the ratios $a_l/a_u$ and $e_l/e_u$ their dependences on $e$ and $a$, respectively, cancel out almost entirely.
As a result, the correlations between \ups\ and the two ratios in Fig.~\ref{fig:upsratiodomega} show only a little scatter of the data around the calculated curves. 

\indent Figure~\ref{fig:upsratiodomega} shows that the ratios $a_l/a_u$ and $e_l/e_u$ get smaller as the orbit gets more relativistic (i.e. \ups~ increases). This means that, for all relativistic orbits, the fit to the upper half has a larger semimajor axis and is more eccentric than the fit to the lower half, as expected. Also, as \ups~goes to zero, both ratios approach unity since less and less deviation from a Newtonian orbit is anticipated. 
The best fits after trying a few models to describe the correlations are

\begin{equation}
{a_l}/{a_u} = \left(-3.14 \pm 0.18\right) {\Upsilon}^{\left(1.15 \pm 0.02\right)} + 1
\label{eqn:upsilonaratio}
\end{equation}

\noindent and

\begin{equation}
{e_l}/{e_u} = \left(-0.41 \pm 0.01\right) {\Upsilon}^{\left(0.44 \pm 0.01\right)} + 1~.
\label{eqn:upsiloneratio}
\end{equation}

\indent Using the periapse distance instead of the relativistic parameter will not give us any new information, since $r_{p} = {r_s} / \Upsilon$.
According to \citet{zuc2006}, the stars with smaller periapse passages and consequently larger velocities at the periapse, i.e. larger $\beta_{p}$, are in orbits with stronger PN effects. Also, for highly eccentric orbits with $r_{p}\ll a$ and $\Upsilon \ll~$1 (approximately Newtonian), \citet{zuc2006} show $\beta \sim \sqrt{\Upsilon}$. Here, we also find $\beta = (0.713 \pm 0.003) \sqrt{\Upsilon}$ as can be seen in Fig.~\ref{fig:upsilonbeta}.

\subsection{The Case of S2}
\label{subsec:S2}

\begin{figure*}[htbp]
\centering
\subfigure{\includegraphics[width=0.32\textwidth]{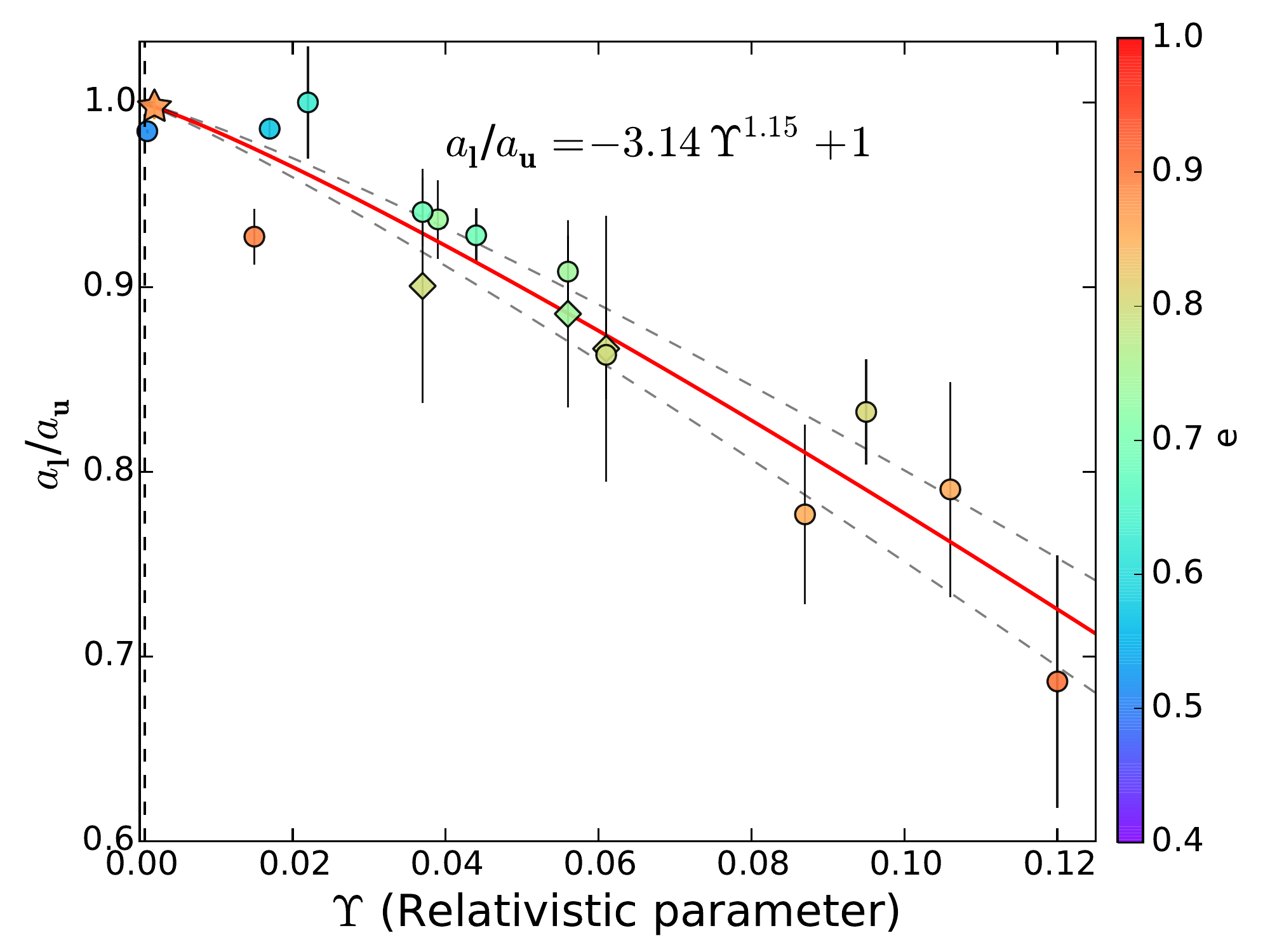}\label{fig:upsilonaratio}}
\subfigure{\includegraphics[width=0.32\textwidth]{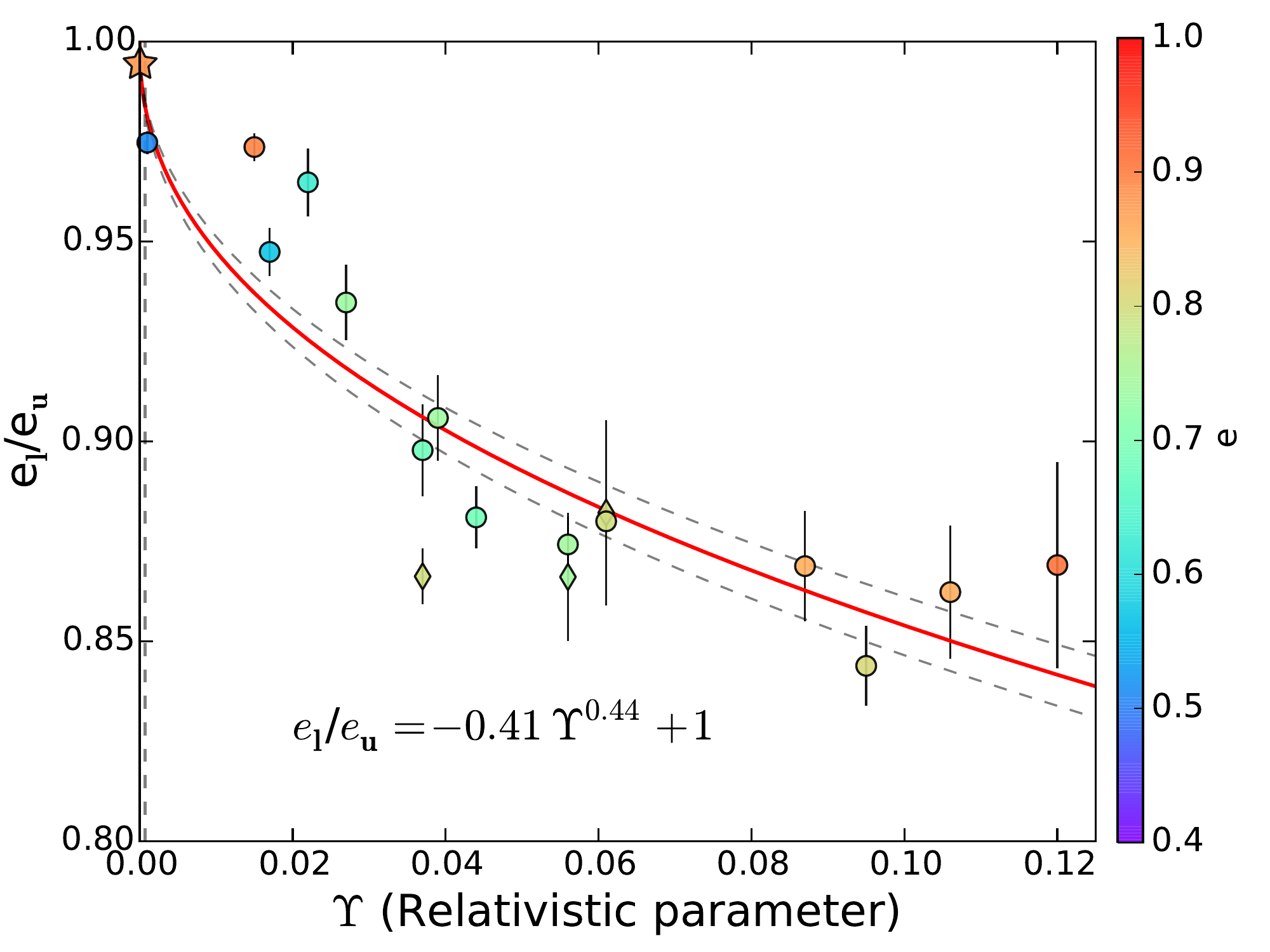}\label{fig:upsiloneratio}}
\subfigure{\includegraphics[width=0.32\textwidth]{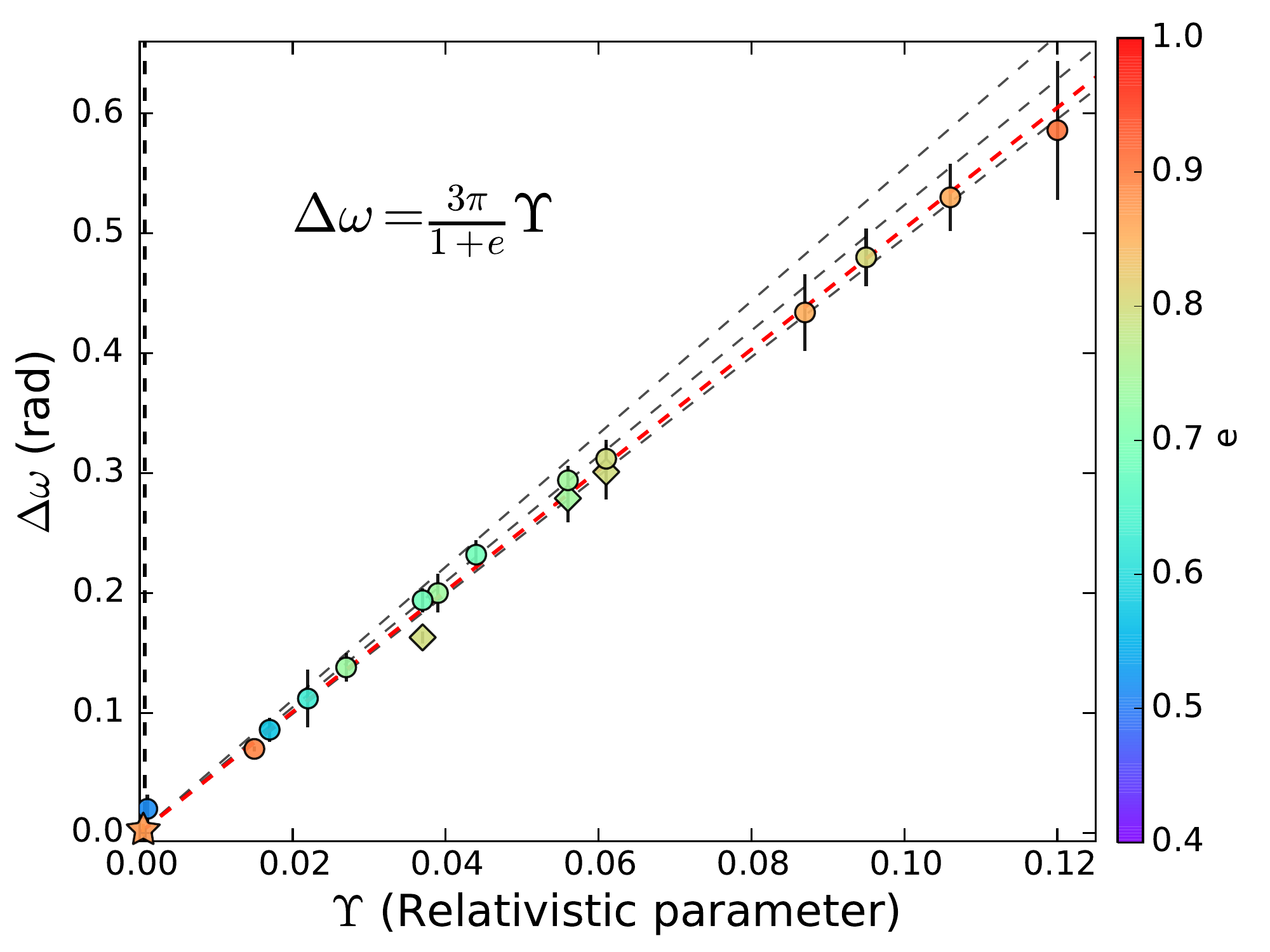}\label{fig:upsilondeltaomega}}
\subfigure{\includegraphics[width=0.32\textwidth]{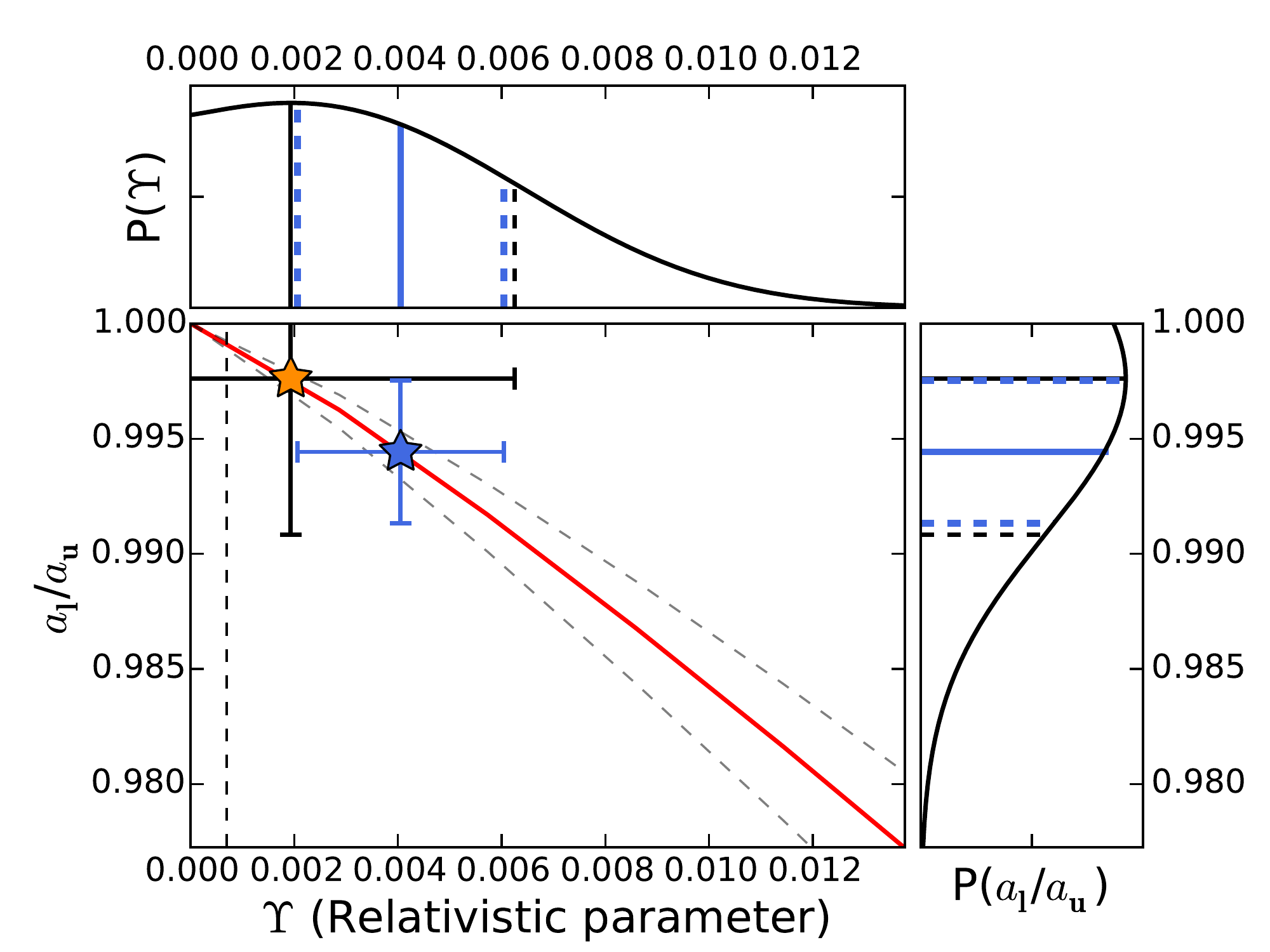}\label{fig:s2aratio}}
\subfigure{\includegraphics[width=0.32\textwidth]{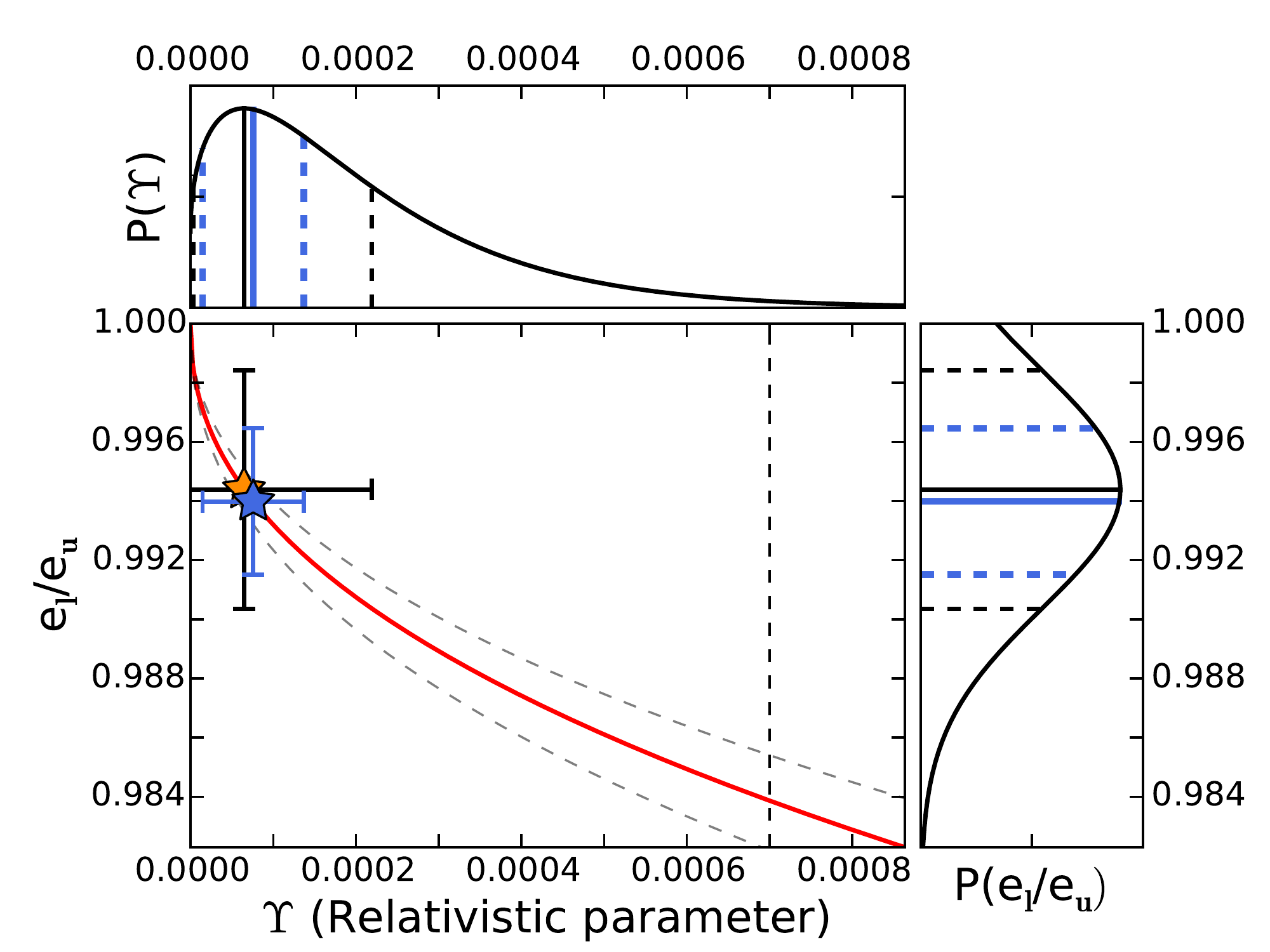}\label{fig:s2eratio}}
\subfigure{\includegraphics[width=0.32\textwidth]{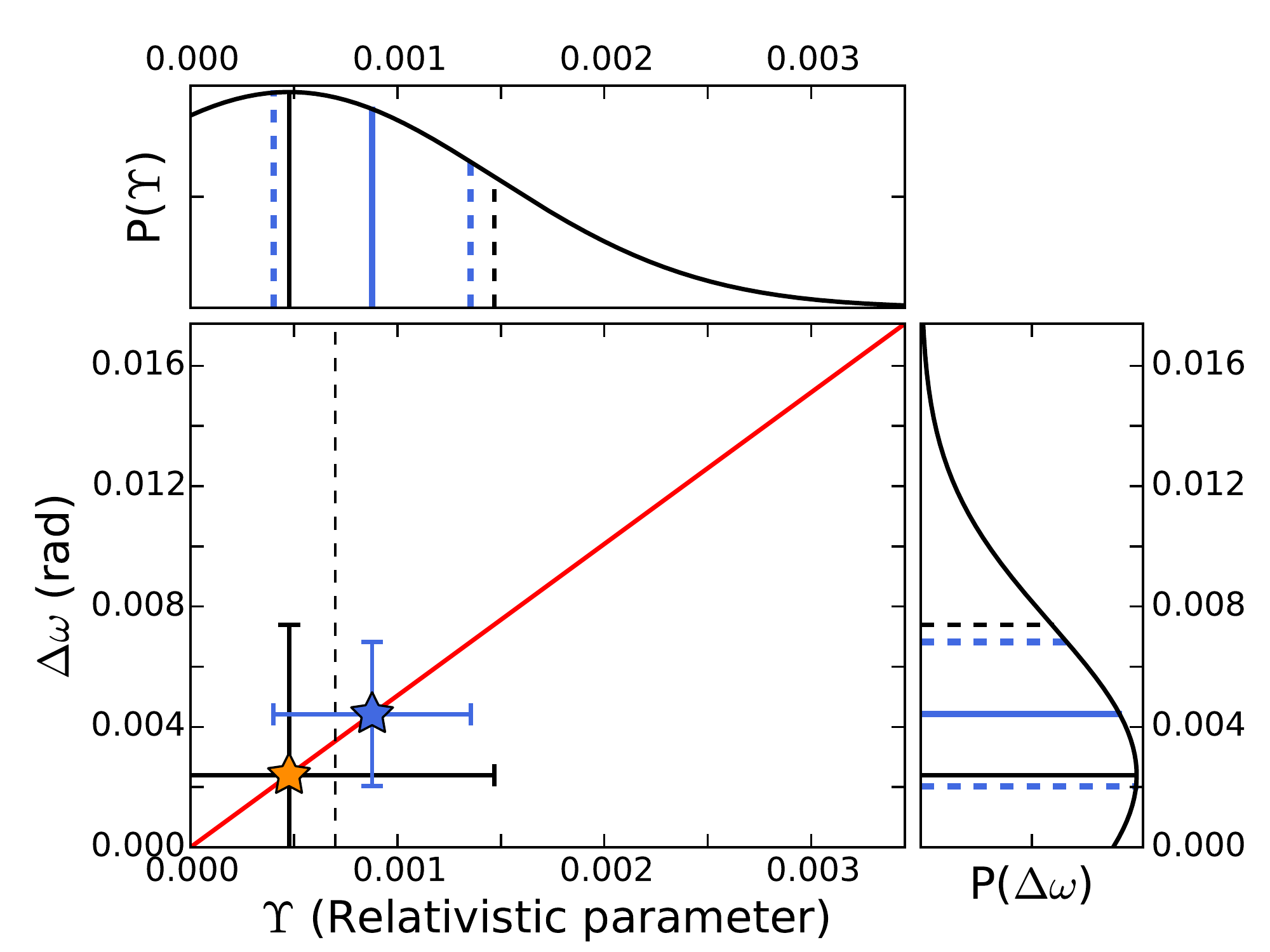}\label{fig:s2deltaomega}}
\caption{{\bf Top panels:} correlation between the relativistic parameter \ups~and 1) the ratios of the orbital elements of the elliptical fits to the lower and upper halves of the orbit---the semimajor axis ($a_l/a_u$) (top left panel) and the eccentricity ($e_l/e_u$) (top middle panel) 2)---and the periapse shift $\Delta \omega$ (top right panel), for the case studies in Table~\ref{table:noi}. 
S2 is represented with a star symbol. The diamonds represent results from inclined orbits (shown corrected for inclination)
and the circles represent results from the orbits without inclination. The color bars show the eccentricities of the orbits. The results for the corresponding orbits with and without inclination are consistent. The correlations are demonstrated with red lines. The gray dashed curves that run alongside the red lines are the uncertainties for the top left and top middle panels. In the top right panel the grey dashed lines (listed from from top to bottom) are the correlation for $e = 0.7$, $e = 0.8$, and $e = 0.9$. The red dashed line is the correlation for $e_{S2} \approx$~0.87. 
The vertical black dashed lines in the top and bottom panels (close to the left edge of the plot in top three panels) represent the expected value of \ups$_{S2}$. The color bars show the eccentricity of the orbits.
{\bf Bottom panels:} here we zoom in to the results of the correlations for S2. The distributions of the $a_l/a_u$, $e_l/e_u$, and $\Delta \omega$ are shown in the small panels next to the $y$-axis. Using the change of variables, the distribution of \ups~is derived and plotted next to the $x$-axis in all three plots. The solid black and dashed lines are the means and the standard deviations, shown as orange stars with error bars, and the solid and dashed blue lines are the medians and the median absolute deviations, shown as blue stars with error bars in all three panels.}
\label{fig:upsratiodomega}
\end{figure*}

\begin{figure}[htbp]
\centering
\includegraphics[width=0.5\textwidth]{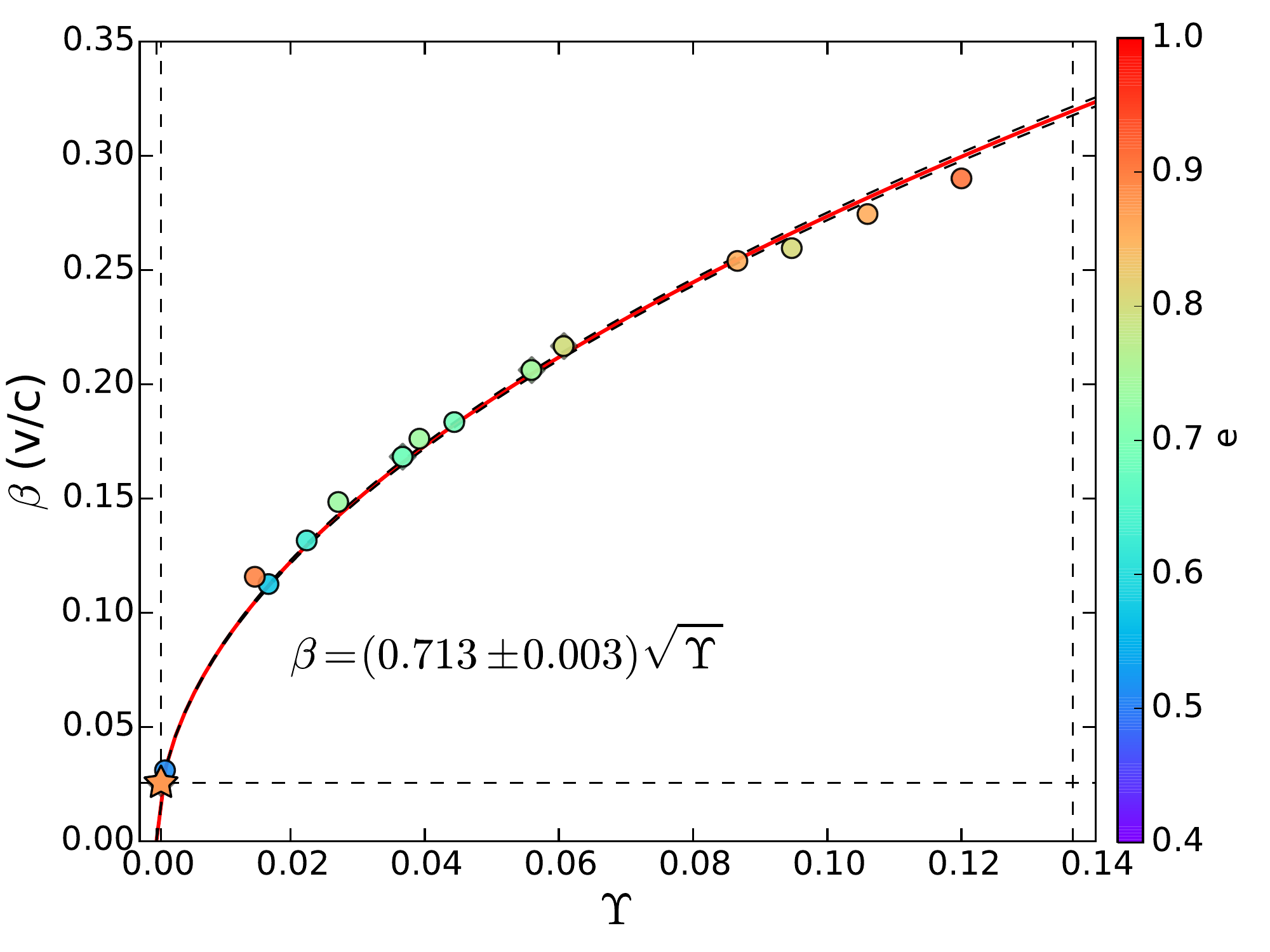}
\caption{Correlation between the relativistic parameter \ups\ and the relativistic $\beta$ at the periapse in units of the velocity of light for the simulated orbits in Table~\ref{table:noi}. The stars with (corrected for) inclination with respect to the sky plane are shown with diamonds and the stars on the sky plane are represented with circles. S2 is depicted by a star symbol. 
The straight dashed lines show limits for orbits chosen in our case study.
These are the velocity of S2 at the periapse at the bottom, \ups$_{S2}$ on the left, and on the right the \ups\ at the periapse passage if the tidal disruption radius is reached for the simulated orbits. The color bar shows the theoretical eccentricity of the orbit of the stars. The correlation is demonstrated with a red line. The black dashed curved lines are the uncertainties of the fit.}
\label{fig:upsilonbeta}
\end{figure}

\indent We also apply the analysis to data on the S-star S2. The orbit of this star shows the highest ellipticity and 
gives us a chance of deriving the relativistic parameter.
The results for $a$ and e from the fits to the upper and lower halves of the combined data set of S2, given in the last row of Table \ref{table:noi}, are the mean and standard deviation of the assumed normal distributions from the MCMC simulations. However, when calculating $a_l/a_u$ and $e_l/e_u$, since both ratios in the derived correlation are limited to 1 (Equation~\eqref{eqn:upsilonaratio} and \eqref{eqn:upsiloneratio}), we choose to use a truncated normal distribution as the probability density function (pdf) given by 

\begin{equation}
f\left(x; \mu, \sigma, a, b\right) = \frac{\left(1/\sigma\right) \phi\left(\left(x-\mu\right)/\sigma\right)}{\Phi\left(\left(b-\mu\right)/\sigma\right) - \Phi\left(\left(a-\mu\right)/\sigma\right)},
\label{eqn:truncpdf}
\end{equation}

for $a \leq x \leq b$ with $\phi$ being the pdf of a standard normal distribution and $\Phi$ being the cumulative distribution function. Using a change of variables and the correlations between the observable parameters and \ups, we can obtain the pdf of \ups. These pdfs are shown next to the $x$- and $y$-axis in the bottom panels of Fig.~\ref{fig:upsratiodomega} for S2. The solid black and dashed lines are the means and standard deviations, respectively.
They correspond to the orange stars in all three panels. Moreover the medians and the median absolute deviations ("mad") are shown with blue solid and dashed lines, respectively. They correspond to the blue stars with their error bars representing the "mad". The median and mean values are listed in Table~\ref{table:s2result}. 

\begin{table*}[htbp]
\centering
\caption{Relativistic parameter of S2 derived from the $a_l/a_u$, $e_l/e_u$ and $\Delta\omega$ methods with and without the drift motion of Sgr A*. 
The individual results are given as means (with standard deviations) and medians (with median absolute deviation). 
In each row the last two columns show the mean value of the \ups\ from the combined $a_l/a_u$ and $e_l/e_u$ methods and the $\Delta\omega$ methods and its standard deviations of the mean, and the median value of the medians and its median absolute deviation.}
\label{table:s2result}
\begin{tabular}{l c c c c c}
\hline\hline
Method & $a_l/a_u$ & $e_l/e_u$ & $\Delta \omega$ & Mean & Median\\
\hline
With the drift\\
motion of BH:\\
\ups~(Mean) & 0.00193 $\pm$ 0.00432 & 0.00006 $\pm$ 0.00015 & 0.00048 $\pm$ 0.00099 & 0.00074 $\pm$ 0.00227 & --\\
\ups~(Median) & 0.00405 $\pm$ 0.00199 & 0.00008 $\pm$ 0.00006 & 0.00088 $\pm$ 0.00048 & $\mathbf{0.00147 \pm 0.00105}$ & $\mathbf{0.00088 \pm 0.00080}$\\
\hline
Without the drift\\
motion of BH:\\
\ups~(Mean) & 0.00179 $\pm$ 0.00424 & 0.00001 $\pm$ 0.00005 & 0.00048 $\pm$ 0.00099 & 0.00069 $\pm$ 0.00223 & --\\
\ups~(Median) & 0.00392 $\pm$ 0.00194 & 0.00002 $\pm$ 0.00002 & 0.00088 $\pm$ 0.00048 & 0.00142 $\pm$ 0.00102 & 0.00088 $\pm$ 0.00086\\
\hline
\end{tabular}
\end{table*}

\indent The strongest cumulative relativistic effect is the shift of the periapse due to a Schwarzschild black hole up to the first order and is given by

\begin{equation}
\Delta\omega = \frac{6\pi G M_{BH}}{c^2a\left(1-e^2\right)}
\label{eqn:perishift_upsilon}
\end{equation}

\noindent per orbit. Hence, for the second method explained in the previous section ($\Delta \omega$ method), we can repeat the approach explained above. Once more the relation (Equation~\eqref{eqn:perishift_upsilon}) is limited only to the positive values for $\Delta \omega$. Therefore to obtain the pdf of \ups~we use the truncated pdf of $\Delta \omega$ as shown in the top left panel of Fig.~\ref{fig:upsratiodomega}. The results are presented in Table~\ref{table:s2result}. Since the values for \ups$_{S2}$ from $a_l/a_u$ and $e_l/e_u$ both describe the case for folding symmetry along the semi-minor axis, we averaged their results. Consequently, the final mean value of \ups$_{S2}$ is obtained by averaging the values corresponding to the orthogonal folding symmetries. The three values obtained for \ups$_{S2}$ are obtained from methods that react on statistical and systematic uncertainties of the data in different ways. The variations that affect $\omega$ act differently on the ellipticity and the semi-major axis. Similarly, the variations that affect the semi-major axis may not result in a change in $\omega$. Hence, the uncertainties between the three different methods certainly do not follow a Gaussian distribution. Therefore, the three values need to be combined by averaging the median and "mad" rather than the mean and standard deviation.

\indent For a full treatment of S2 the drift motion of the central mass needs to be taken into account. This motion should be added to the coordinates of the ellipse we are fitting at each time step. Alternatively, the motion can be removed from the data points. If we consider the motion from Table \ref{table:allresults} for the relativistic fit to only S2 data and remove it from the combined data set of S2, we get similar results as can be seen in Table \ref{table:s2result}.

\subsection{Robustness of the Result}
\label{robustness}

\indent We can address the robustness of the result in different ways. 
First we exclude the possibility that the result is dominated by noise or by a drift motion of Sgr~A*, then we highlight again that it is not due to rotation 
between the VLT and Keck data sets.

\indent We assume that the orbital measurements are completely dominated by the noise and that the signal 
showing the variations of the orbital elements is dominated by noise. For simplicity we consider a displacement 
of an orbital section from the noise-unaffected orbit due to noise as a single degree of freedom.
One can ask the question: how likely is it to get consistent results as presented in this section assuming that the noise contributions of the upper and lower halves and the pre- and post-periapse halves of the orbit are independent? Furthermore, we consider the collective noise contribution in each one of these quadrants as a displacement along the semimajor or semiminor axis of a single orbit. While no significant net displacement will occur for most of the noise realizations, a certain fraction of these realizations will result in such a displacement. Hence, the probabilities derived below are crude lower limits if one seeks to explain the result as being due to the noise only. Considering only the fraction of the noise realizations that result in a net displacement we have a total of 4$^4$ = 256 possibilities
to combine them. Only one combination gives a unique configuration that results in the measured changes of $\omega$, and the ratios $a_l/a_u$ and
$e_l/e_u$. This corresponds to a probability of 0.004, i.e., a 0.4\% chance that the result is reproduced by noise-dominated measurements. 

\indent Allowing for at least one quadrant to be displaced improperly leaves us with five combinations that are consistent with the observed signal. Allowing for more than one of the four quadrants to be displaced improperly gives an inconsistent result for at least one of the quantities, $\omega$, $a_l/a_u$, or $e_l/e_u$. Hence, there is only a 2\% probability (5/256 = 0.019) that the result can be obtained serendipitously as a consequence of dominant noise. Therefore, we consider it to be highly unlikely that the common systematic tendency of the change in the orbital parameters and hence the corresponding ranges for the relativistic parameter are the result of pure noise only.

\indent It can also be argued that the change in the orbital parameters is due to the drift motion of Sgr~A*. In this case a constant shift in $\omega$ would be injected over the entire orbital time scale. For one orbit the drift would be of the order of $a$$\Delta$$\omega$, with $a$ being the semimajor axis of the orbit and $\Delta \omega$ the expected periastron shift. There would be no effect on $\omega$ for a north--south motion. However, in east--west direction a proper motion of Sgr~A* of the order of 30~$\mu$as~yr$^{-1}$ could explain the shift in $\omega$. Nevertheless, by using the PN approximation for modeling the motion of S2 and correcting the orbital data for the residual drift motion before deriving the change in the orbital parameters, we can assume that it is very unlikely that the observed change in sign and magnitude of $\omega$ is due to the drift motion of Sgr~A*.

\indent It can also be excluded that the change in the orbital parameters is due to a relative rotation between the data sampling the pre- and post-periapse halves of the orbit. Both data sets are tied to the same VLA radio reference frame. Moreover the classical calibration of the camera rotation is better than 0$\degr$.1, i.e. less than 6$'$, and therefore about a factor of two smaller than the expected periastron shift of about 11$'$ (for a semimajor axis of 0\arcsec.126, an eccentricity of 0.88, and a BH mass of 4.15~$\times$~10$^6$~\msun). Furthermore,
\citet{plewa2015} find an upper limit on the temporal rotation $(v_{\phi}/r)$ of the infrared reference frame relative to the 
radio system of $\sim$7.0~$\mu$as~yr$^{−-1}$~arcsec$^{−-1}$. Over 20 yr this corresponds to an angle of less than 0$'$.5, 
i.e. 24 times smaller that the expected periastron shift. 
This is also consistent with our result from section~\ref{sec:rels2}.
A comparison of the VLT and Keck data at times of equal coverage 
for S38 or for the combination of 
S2 and S38 results in an upper limit of the rotation 
of 0$'$.1, i.e., a value 110 times smaller than the expected periastron shift.
Therefore, we consider it very unlikely that the observed change in sign and magnitude of $\omega$ is due to the effects of differential rotation.

\section{Discussion}
\label{sec:comb}

\subsection{Comparison of the Results with the Literature}
\label{sec:complit}
\indent The next periapse passage of S2, assuming the values from simulations of all three stars with Newtonian models, is in 2018.51~$\pm$~0.22, which is in July. \citet{boehle16} have predicted it to be 2018.267~$\pm$~0.04, corresponding to April.
In both cases the results indicate that the event might be optimally placed for observations. 
 The upcoming event is highly anticipated since the deviations from a Newtonian orbit and the gravitational redshift are expected to be detectable as S2 goes through its closest approach. However, these tests of GR are possible only if we have a precise knowledge of the gravitational potential parameters and the orbit. Using the data from more than one star for orbital fitting is one way of getting a better precision in finding these parameters. Using multiple stars for determining M$_{BH}$ and R$_0$ has been done before. \citet{gill17} find $M_{BH}$~=~(4.28~$\pm$~0.10)~$\times$~10$^6$~\msun\ and $R_0$~=~8.32~$\pm$~0.07~kpc for multiple-star fit. While the statistical uncertainties of these parameters are comparable to the uncertainties we report in our Newtonian multiple-star fit, the results reported by \citet{gill17} are in agreement with our values to within a 2$\sigma$ uncertainty (see Table~\ref{table:allresults}).
Similarly, \citet{boehle16} measurements from Keck are $M_{BH}$~=~(4.02~$\pm$~0.16)~$\times$~10$^6$~\msun\ and $R_0$~=~7.86~$\pm$~0.14~kpc. To within a 1$\sigma$ and $\sim$2$\sigma$ uncertainties, respectively, these quantities are in agreement with our comparable fit results (Newtonian S2 and S38). The uncertainties we obtain are very similar to those of \citet{boehle16}.

\indent The star S55/S0-102 was not selected before for mass and distance fits due to the lack of radial velocity data. Since we cannot constrain the Newtonian precession due to the uncertainty of the mass enclosed within the orbit of S2, we need at least one more star to measure the strength of the PN effects. S55/S0-102 has a very short orbital period and thus a large phase coverage. 
It has already passed through its periapse passage in 2009 and its next periapse time will be in 2021. Therefore, it is the best candidate for measuring deviations from a Newtonian orbit after S2. 
The parameters we derive for the Newtonian model using the combinations S2, S38, and S2, S38, S55/S0-102 give similar results within the uncertainties.

\indent The estimates of mass and distance derived for the Newtonian and the relativistic models and being based solely on S2 give slightly higher results 
that are, however, within their larger uncertainties, still in agreement with the multiple-star solutions. They also agree very well with recent estimates of these quantities by \cite{ghez8} and \cite{gill2009} based on Newtonian solutions for S2 only.

\indent Also the estimates of mass and distance for the S2, S38 combination in the relativistic case are in reasonable agreement with the Newtonian solutions,
while these quantities are systematically larger in the relativistic case for the S2, S38, S55/S0-102 solution.

\indent The ratios of the changes in orbital parameters of S2---$a_l/a_u$ and $e_l/e_u$---and $\Delta \omega$ given in Table~\ref{table:noi} agree with the study done by \citet{iorio17} where time series of the first-order PN shifts of the osculating Keplerian orbital elements are found analytically and numerically. \citet{iorio17} finds maximum shifts of $\Delta a$~=~30~au, $\Delta e$~=~0.003, and $\Delta \omega$~=~0$\degr$.2 for S2. Considering a semimajor axis of $a$ = 0\arcsec.126 and an eccentricity of $e$ = 0.884, we get $a_l/a_u = (a - \Delta)/a$ = 0.971 and $e_l/e_u = (e - \Delta)/e$ = 0.997, which are in agreement with the values in Table \ref{table:noi}. The results described by \citet{iorio17} were obtained in a deductive way: a general relativistic theoretical scenario is used to deduce the orbital elements and their properties as a function of time, as one would expect them for the special star S2 close to the Sgr~A* black hole. In our case we have to proceed in an inductive way. We start with positional and spectroscopic measurements of the three stars (including S2) that orbit the large mass associated with Sgr~A*. The goal is to show that the orbit of S2 is significantly influenced by general relativistic effects. Hence, we have to provide an (indirect) observable that allows us to discriminate the relativistic from the nonrelativistic case, based on real data. The procedure we developed can be directly compared to and confirms the predictions given by \citet{iorio17}.

\subsection{Overcoming the Bias in the Orbital Fitting}
\label{sec:morestar}

\indent While the general agreement between the Newtonian and relativistic fits is good, it appears that the fits still need 
to be further constrained. 
While a Newtonian model seems to be able to describe the trajectory of all S-stars so far, we have to use a relativistic model if we expect the orbit to show any precession at the periapse. This is due to the fact that when fitting a Newtonian model, a small precession can be compensated by a larger drift motion of the SMBH in the same direction of the precession if we have observed the star for only about one orbital period. Even with a relativistic model, we need at least two stars with preferably different orientations on the sky (such as S2 and S38) to be able to overcome this bias.

This may be achievable in the near future by including more stars.
However, using more stars that are at larger distances from Sgr~A* would only bias it towards a Newtonian solution.
What is needed is the inclusion of more stars closer to---or at a similar distance from---Sgr~A* than the currently used trio.
Observations at higher angular resolution with a high point-source sensitivity, e.g., with GRAVITY at the VLTI
\citep{Eisenhauer2011, Eckart2012, Grould2017} or cameras at telescopes such as the E-ELT \citep{Brandl2016, Davies2016}, will help in reaching this goal. 

\subsection{Detectability of the PN Effects}
\label{sec:observpn}
\indent Although most of the deviations from a Newtonian orbit happen during the closest approach to the SMBH, the measurement is not an easy task
since the IR counterpart of the SMBH may be confused with other sources during the periapse passage. 
The correlation between \ups\ and the ratios $a_l/a_u$ and $e_l/e_u$ (as discussed in Sect.~\ref{sec:sim}), can provide us with an estimate of \ups\ after observing the star for one orbit, which can consequently result in the prediction of $r_p$, $\beta_p$, and $\Delta \omega$. All these values cannot be measured directly without knowledge of the orbit. 

\cite{AngelilSaha2014} propose to measure $\Delta \omega$ in one full orbit since $\omega$ at each instant is not observable and one cannot simply measure it before and after the periapse. We measure this parameter by fitting elliptical orbits to the entire S2 data set before and after periapse and derive \ups\ by utilizing the correlation between $\Delta \omega$ and \ups. We find the median of the resulting medians from the three \ups\ distributions and its median absolute deviation to be \ups~=~0.00088~$\pm$~0.00080. Alternatively, since the $a_l/a_u$ and $e_l/e_u$ methods use the same symmetry in the orbit we take the average of the medians of the \ups\ distributions, i.e., the one derived from the correlations between \ups\ and the ratios $a_l/a_u$ and $e_l/e_u$ considering them as one method by using their mean value, and the \ups\ value from $\Delta \omega$ method. We find \ups~=~0.00147~$\pm$~0.00105. Both approaches result in values that are consistent within the uncertainties with the expected value of 0.00065 for S2 (for a semimajor axis of 0\arcsec.126, an eccentricity of 0.88, and a BH mass of 4.15~$\times$~10$^6$~\msun). The use of the medians of the distributions instead of their means for the combination of the results is justified in Sect.~\ref{subsec:S2} One might argue that the drift motion of the SMBH might have affected our result. A large enough north--south motion will affect the semimajor axis of the orbit and an east--west motion will increase or decrease the periapse shift. To investigate this, we apply the correlations to the drift motion corrected combined data of S2 and obtain the average value of 0.00069~$\pm$~0.00223,
which is even closer to the theoretically predicted value. The predicted drift motion is taken from Table~\ref{table:allresults} for the relativistic best fit of S2. This means that the drift motion of Sgr~A* does not affect our result significantly, and for S2 the combination of these methods can successfully give a measure of the PN effects without going through complex relativistic fitting procedures. Our method is limited by the fact that the data for the pre-periapse and the bottom half of the orbit are sparser and more uncertain before 2002
than the remainder of the data. This limitation will be resolved only after S2 has reached another apoapse in 2026.

\indent Moreover, if we consider the expected value of 0.00065 for the \ups\ of S2, using the correlation between \ups\ and the relativistic $\beta$ from Sect.~\ref{subsec:casestudies}, we find $\beta_{S2} =$~0.001818~$\pm$~0.00008. If we use \ups\ of S2 derived in this work (0.00069$\pm$ 0.00223) we find $\beta_{S2} =$~0.001873~$\pm$~0.03027. Both values agree with $\beta_{S2} \sim$~0.02 from the simulations of the orbit of S2.

\section{Conclusion}
\label{sec:con}

\indent In order to derive the mass of the SMBH Sgr~A* and the distance to the GC, we used the three stars 
S2, S38, and S55/S0-102, which are currently known to be the closest to the center. We find $M_{BH} = 4.15 \pm 0.13 \times 10^{6}$~\msun\ and $R_0 = 8.19 \pm 0.11$~kpc based on Newtonian orbital models, which are in good agreement with the recently published values. There are systematic errors on these values due to the possibility of choosing a relativistic model instead of a Newtonian one \citep{gill17}. The differences in $M_{BH}$ and $R_0$ between Newtonian and relativistic models (Table~\ref{table:allresults}) are 0.57~$\times 10^{6}$~\msun\ and 0.34~kpc, respectively. We expect that the errors unaccounted for in the construction of the reference frame (some of it is accounted for by including the standard deviation of the mean of the residuals of the five reference stars in the astrometric errors of the S2, S38, and S0-102/S55 described in Sect.~\ref{sec:orbitsderiv}) are small in comparison to these values (0.04 $\times 10^{6}$ \msun~and 0.04 kpc in the calculation done in \citet{boehle16}. Hence the systematic errors are dominated by the differences in model, and our final best estimates are $M_{BH} = 4.15 \pm 0.13 \pm 0.57 \times 10^{6}$~\msun\ and $R_0 = 8.19 \pm 0.11 \pm 0.34$~kpc.

\indent We use the first-order PN approximation to simulate the relativistic orbits for a broad range 
of the impact parameters. We present two methods that utilize the changes in
the orbital parameters to measure the relativistic parameter at the closest approach to Sgr~A*. 
The results from these methods determined for the simulated orbits are then applied to the orbital analysis of S2. Consequently, we are able to determine a consistent change in the orbital elements of S2 from the differences between the orbital fits to the upper/lower and pre-/post-periapse parts of the orbit. These changes imply a relativistic parameter of \ups~=~0.00088~$\pm$~0.00080 which is within the uncertainties consistent with the expected theoretical value of \ups~=~0.00065 for the star S2 orbiting Sgr~A* (for a semi-major axis of 0\arcsec.12600, an eccentricity of 0.88, and a BH mass of 4.15~$\times$~10$^6$~\msun). For the changes in the argument of periapse we find a median with median absolute deviation of $\Delta \omega = 14 \pm 7$\arcmin\ (or $\Delta \omega = 14 \pm 13$\arcmin\ applying the range for \ups\ derived above) which is consistent with 11\arcmin, expected for S2 lowest order periapse shift. Since the eccentricity of S2 is the largest among the three stars, it is currently the star best suited for the determination of the relativistic parameter.

\indent This result must be seen in the light of the discussion of the resonant relaxation within the central 
star cluster close to Sgr~A* \citep{rauch96, alex2005, hopman2006, merritt10, Koc&Trem11, sabha12}.
\citet{sabha12} investigate the effects of the orbital
torques on the orbit of S2 due to the resonant relaxation. They
find that if a significant population of 10\msun\ black holes is present, which account for an enclosed mass
between 10$^3$~\msun\ and 10$^5$~\msun\ \citep[e.g. see][]{Mouawad2005, Freitag2006}, then the contributions from
the scattering will be important for the trajectory of S2. The
authors show that the effect for each single orbit can
be of the same order of magnitude as the relativistic or
Newtonian periapse shifts. The scatter of this effect is
large and can lead to a significant apparent weakening
or enhancement of the relativistic shift (see Figs. 9 and 10 in \citet{sabha12} and the description of the cases in their section 4). Also the effect may be different from
one orbital period to the next. Hence, the additional contributions to the relativistic shift of the
order of 10$'$ or even more would indicate that a significant 
population of massive objects (each a few tens of solar masses)
influences the orbit of S2.

\indent Taking the derived relativistic parameter of \ups\ at face value implies that, at least over 
the orbital time scale of the S2 resonant relaxation, the proper motion of Sgr~A* within the stellar cluster as well as the effect of
an extended mass are not relevant within the current measurement uncertainties.

\indent If one argues that the derived changes in the orbital parameters of S2 are random results or dominated by
the disturbing effects discussed by \citet{sabha12}, then one must claim that all these effects
compensate each other such that in the sign and magnitude the theoretically predicted value of the
relativistic parameter is obtained to within the 1$\sigma$ uncertainties.

\indent In future, continued single-dish or interferometric
studies of the stellar orbits close to Sgr~A* 
must be performed in order to determine the relativistic parameter of other stars and to 
further control the influence of the drift motion of Sgr~A*.

\acknowledgments

The authors would like to thank the referee for her/his constructive suggestions. We received funding6
from the European Union Seventh Framework Program
(FP7/2013-2017) under grant agreement no 312789 - Strong
gravity: Probing Strong Gravity by Black Holes Across the
Range of Masses.
This work was supported
in part by the Deutsche Forschungsgemeinschaft (DFG)
via the Cologne Bonn Graduate School (BCGS) and the SFB 956 - Conditions and Impact of Star Formation, the
Max Planck Society through the International Max
Planck Research School (IMPRS) for Astronomy and
Astrophysics, as well as special funds through the
University of Cologne and
the SFB 956; Conditions and Impact of Star Formation. 
M. Parsa and M. Zaja\v{c}cek are members of the IMPRS.
Part of this
work was supported by fruitful discussions with members of
the European Union funded COST Action MP0905: Black
Holes in a Violent Universe and the Czech Science Foundation
- DFG collaboration (No. 13-00070J).

\bibliography{mybib}{}
\bibliographystyle{aasjournal}

\end{document}